\documentclass[aps,prd,twocolumn,showpacs,notitlepage,eqsecnum,
superscriptaddress]{revtex4-1}

\usepackage[centertags]{amsmath}
\usepackage{amssymb}
\usepackage{latexsym}
\usepackage{enumerate}
\usepackage{graphicx}
\usepackage{mathrsfs}
\usepackage{hyperref}
\usepackage{stmaryrd}
\usepackage{import}
\usepackage{tensor}
\usepackage{color}
\usepackage{bm}
\usepackage{multirow}
\usepackage{breakurl}
\usepackage{float}
\usepackage{placeins}
\usepackage{url}

\newcommand{\bi}{\begin{itemize}}
\newcommand{\ei}{\end{itemize}}
\newcommand{\be}{\begin{equation}}
\newcommand{\ee}{\end{equation}}

\renewcommand{\l}{\left(}
\renewcommand{\r}{\right)}
\renewcommand{\a}{\alpha}

\newcommand{\g}{\gamma}
\newcommand{\G}{\Gamma}

\newcommand{\D}{\Delta}
\newcommand{\e}{\epsilon}

\renewcommand{\O}{\Omega}
\renewcommand{\o}{\omega}

\newcommand{\q}{\quad}
\newcommand{\qq}{\qquad}

\newcommand{\vp}{\varphi}

\allowdisplaybreaks

\begin{document}

\title{Tidal heating and torquing of the primary black hole in eccentric-orbit, non-spinning extreme-mass-ratio 
inspirals to 22PN order}

\author{Christopher Munna}
\affiliation{Department of Physics and Astronomy, University of North Carolina, Chapel Hill, North Carolina 27599, USA}
\affiliation{MIT Kavli Institute, Massachusetts Institute of Technology, Cambridge, MA 02139, USA }
\author{Charles R. Evans}
\affiliation{Department of Physics and Astronomy, University of North Carolina, Chapel Hill, North Carolina 27599, USA}
\author{Erik Forseth}
\affiliation{Graham Capital Management, Rowayton, CT 06853, USA}

\begin{abstract}
We calculate the high-order post-Newtonian (PN) expansion of the energy and angular momentum fluxes onto the horizon 
of a nonspinning black hole primary in eccentric-orbit extreme-mass-ratio inspirals.  The first-order black 
hole perturbation theory calculation uses \textsc{Mathematica} and makes an analytic expansion of the 
Regge-Wheeler-Zerilli functions using the Mano-Suzuki-Takasugi formalism.  The horizon absorption, or tidal heating 
and torquing, is calculated to 18PN relative to the leading horizon flux (i.e., 22PN order relative to the 
leading quadrupole flux at infinity).  Each PN term is a function of eccentricity $e$ and is calculated as a series 
to $e^{10}$.  A second expansion, to 10PN horizon-relative order (or 14PN relative to the flux at infinity), is 
computed deeper in eccentricity to $e^{20}$.  A number 
of resummed closed-form functions are found for the low PN terms in the series.  Using a separate Teukolsky 
perturbation code, numerical comparisons are made to test how accurate the PN expansion is when extended to a 
close $p=10$ orbit.  We find that the horizon absorption expansion is not as convergent as a previously computed 
infinity-side flux expansion.  However, given that the horizon absorption is suppressed by 4PN, useful results can 
be obtained even with an orbit as tight as this for $e \le 1/2$.  Combining the present results with our earlier 
expansion of the fluxes to infinity makes the knowledge of the total dissipation known to 19PN for eccentric-orbit 
nonspinning EMRIs. 
\end{abstract}


\maketitle

\section{Introduction}
\label{sec:intro}

In the past eight years nearly one hundred compact-binary mergers have been observed as gravitational-wave 
events with LIGO and Virgo \cite{AbboETC21a}.  When the launch of LISA occurs \cite{LISA} we anticipate added 
gravitational-wave discoveries, including extreme-mass-ratio inspirals (EMRIs).  Theoretical 
modeling of EMRIs is important for both source detection and parameter estimation 
\cite{BabaETC17,AmarETC17,BaraETC18,BerrETC19}.  For EMRIs, the small mass ratio $\epsilon =\mu/M \ll 1$ 
(e.g., secondary mass $\mu \sim 10 M_\odot$ and primary mass $M\sim 10^6 M_\odot$) serves as an expansion parameter 
and the theoretical calculation utilizes black hole perturbation theory (BHPT) and gravitational 
self-force \cite{PoisPounVega11} techniques.  Example calculations of full EMRI inspirals can be found in 
\cite{OsbuWarbEvan16,WarbOsbuEvan17,MccaOsbuBurt21,LyncVandWarb22}.  In the early inspiral, when the orbit is wide 
(irrespective of mass ratio), post-Newtonian (PN) theory \cite{Blan14} can be applied.  When both limits pertain, 
self-force quantities (e.g., gravitational-wave fluxes and conservative sector gauge invariants) can be found as 
analytic PN expansions.  Early examples include plucking off \cite{CutlETC93} the apparent (later verified 
\cite{Wise93,Pois93}) analytic coefficient in the 1.5PN tail in the energy flux using a numerical BHPT calculation 
and the leading few terms in the redshift invariant for circular-orbit EMRIs \cite{Detw08}.

This paper addresses the horizon absorption of energy and angular momentum in eccentric-orbit EMRIs onto a 
nonspinning (Schwarzschild) primary.  The horizon fluxes are given in analytic form as a simultaneous high-order 
PN expansion and expansion in powers of eccentricity $e$.  This work is the latest in a sequence of papers that 
have made similar expansions of other physical quantities in eccentric-orbit EMRIs in the PN limit using symbolic or 
extremely high precision numerical BHPT calculations.  Initially, Forseth, Evans, and Hopper \cite{ForsEvanHopp16} 
used a numerical implementation of the MST (Mano-Suzuki-Takasugi) method \cite{ManoSuzuTaka96a,ManoSuzuTaka96b} 
in \textsc{Mathematica} to solve the Regge-Wheeler-Zerilli (RWZ) equations \cite{ReggWhee57,Zeri70} to extremely 
high numerical precision (e.g., 200 digits).  Energy fluxes at infinity were calculated on a two-dimensional grid 
of orbits over separation $p$ and eccentricity $e$.  These data were fit to an understood general form of the PN 
expansion, generating numerical coefficients through 7PN order.  The underlying analytic form (e.g., rational numbers 
or rationals times transcendental numbers) of the coefficients were then determined using PSLQ, an integer relation 
algorithm \cite{Fors16}.  A following paper \cite{MunnETC20} made significant improvement to the method by fitting 
individual $lmn$ modes separately, allowing energy and angular momentum fluxes to be found to 9PN and high order 
in $e$.  A third paper \cite{Munn20} extended those results via a fully symbolic approach (originally developed 
in \cite{BiniDamo13,BiniDamo14a,BiniDamo14b,BiniDamo14c,KavaOtteWard15,HoppKavaOtte16}), reaching 19PN at 
$e^{10}$ and 10PN at $e^{20}$.  The depth of the expansion in $e$ revealed closed-form functions in eccentricity 
at numerous PN orders, and led to an understanding of the form of certain logarithmic PN sequences to arbitrarily 
high PN order \cite{MunnEvan19a,MunnEvan20a}.  Lately, we have applied these techniques to the conservative sector, 
computing the redshift invariant to 10PN and $e^{20}$ \cite{MunnEvan22a} and the spin-precession invariant to 9PN 
and $e^{16}$ \cite{MunnEvan22b}, with additional PN terms found to yield closed-form functions of $e$.

Our calculation of the horizon absorption applies the same techniques but looks instead at the asymptotic behavior 
of the downgoing wave at the horizon.  To understand the PN depth of our calculation, we first recall the relative 
magnitudes of leading-order fluxes at infinity and the horizon.  The leading quadrupole energy flux at infinity 
scales with the fifth power of the PN compactness parameter (i.e., $1/p^5$).  Higher multipoles, corrections, and 
tail effects appear at still higher integer and half-integer PN orders.  It is typical to refer to higher-order terms 
by their PN order \emph{relative} to the dominant infinity-side quadrupole flux.  For example, when we say we 
computed the infinity-side flux to 19PN, that would be a term scaling as $1/p^{24}$.  On the horizon side, 
early work on circular orbits by Gal'tsov \cite{Galt82} (for nonspinning and spinning primary), Poisson and Sasaki 
\cite{PoisSasa95} (nonspinning), and Tagoshi, Mano, and Takasugi \cite{TagoManoTaka97} (spinning), showed that the 
leading flux is suppressed by 4PN relative to the infinity-side flux in the Schwarzschild case and by 2.5PN 
relative in the Kerr case.  Our calculations in this paper are restricted to the Schwarzschild case.  With this 
preface, in this paper we give the horizon fluxes in analytic form to 14PN (in an $e^{20}$ eccentricity expansion) 
and to 22PN (in an $e^{10}$ expansion) relative to the leading infinity-side flux.  Alternatively, these results 
can be thought of as 10PN (in the $e^{20}$ expansion) and 18PN (in the $e^{10}$ expansion) relative to 
the leading horizon flux.

Most of our results come from use of the fully symbolic approach \cite{Munn20}, but some of the lowest-order terms 
were found by Forseth by fitting \cite{Fors16} and then tantalizingly resummed into closed-form expressions in $e$ 
(see also \cite{Fors16b,Evan16}).  Earlier, Shah \cite{Shah14} had used high-precision fitting to find the 
circular-orbit horizon flux expansion past 20PN in a mixed numeric-analytic form.  Then, Fujita \cite{Fuji15} 
derived an entirely analytic expansion for the horizon energy flux for circular orbits about a nonrotating 
black hole to 22.5PN relative order (i.e., to 18.5PN order relative to the leading horizon term).  For the case of 
a Kerr primary, Fujita \cite{Fuji15} computed the expansion to 11PN relative (8.5PN relative to the leading Kerr 
horizon flux).  The effects of eccentricity and inclination were then found \cite{SagoFuji15} to 4PN relative 
(1.5PN past leading horizon term) and $e^6$.  This result was later extended to 5PN relative (2.5PN past leading 
horizon term) and $e^{10}$ in \cite{FujiShib20,IsoyETC22}.  Reducing our results to the circular-orbit limit, 
we match Fujita \cite{Fuji15} completely to 22PN relative to the dominant flux to infinity.  Recently, horizon 
fluxes (tidal heating) have been discussed as a means to distinguish black holes from exotic compact objects in 
coalescing binaries \cite{DattETC20}.  See \cite{ChatPoisYune13} for a calculation of tidal heating and torquing 
in a generic binary encounter and \cite{ChatPoisYune16} for the case of a quasicircular orbit.  

The depth of our calculations precludes us publishing here the full PN expansions.  Instead, we detail in this paper 
the form of the eccentricity dependence of each term through 8PN relative to the leading horizon flux (12PN relative 
to the full flux).  The full expansions are posted online \cite{UNCGrav22}, including on the Black Hole 
Perturbation Toolkit website \cite{BHPTK18}.  The results displayed here are sufficient to note similarities with 
the infinity-side fluxes discussed previously \cite{MunnEvan19a,MunnETC20,MunnEvan20a}.  Using the full 
expressions, we evaluate the 
PN expansions numerically at a separation of $p = 10$ and a set of different eccentricities and compare those 
values to accurate horizon fluxes derived from a full BHPT code.  This is analogous to the computation done in 
\cite{Munn20} to determine the fidelity and convergence of the PN expansions of the infinity-side fluxes.  A set of 
different resummations of the series is examined.  We find the series exhibit useful convergence at $p = 10$ for 
low eccentricity orbits ($e = 0.01$ and $e = 0.1$) but cease to converge for an orbit this tight beyond a few 
orders of magnitude when the eccentricity reaches $e = 0.25$ and is essentially useless at $e = 0.5$.  At $e = 0.25$ 
the best fractional error reaches $10^{-3}$ or slightly better, which can still be useful for inspiral simulations 
since at this orbital separation the horizon flux will be suppressed by four orders of magnitude compared to the 
infinity-side flux.  

The structure of this paper is as follows.  Sec.~\ref{sec:MSTreview} briefly outlines the notation and formalism used
for our analytic expansion procedure.  In Sec.~\ref{sec:EfluxExps} we present the PN and $e$ expansions of the 
energy flux at the horizon, making note of which terms are completely known in $e$ dependence (closed forms) and 
which are known in accurate power series.  Sec.~\ref{sec:LfluxExps} gives the equivalent expansion of the 
angular momentum flux at the horizon.  Sec.~\ref{sec:Disc} gives a general discussion of both the energy and 
angular momentum flux results.  Following that, Sec.~\ref{sec:numComps} presents comparisons to numerical 
flux data to test the validity of the PN expansion.

Throughout this paper we adopt units such that $c = G = 1$, though $\eta = 1/c$ is briefly reintroduced as a 
PN-expansion bookkeeping device.  We use metric signature $(-+++)$.  Our notation for the RWZ formalism follows 
that found in \cite{ForsEvanHopp16, MunnETC20}, which in part derives from notational changes for tensor spherical 
harmonics and perturbation amplitudes made by Martel and Poisson \cite{MartPois05}.  For the MST formalism, we 
largely follow the discussion and notation found in the review by Sasaki and Tagoshi \cite{SasaTago03}.

\section{Brief review of RWZ and MST formalisms}
\label{sec:MSTreview}

Our formalism for solving the first-order black hole perturbation problem for eccentric-orbit EMRIs on a 
Schwarzschild background has been detailed previously \cite{HoppEvan10,ForsEvanHopp16,Munn20}, including the 
added requirements in obtaining fully analytic forms for the PN expansions \cite{Munn20}, which is based on earlier 
work in \cite{BiniDamo13,BiniDamo14a,BiniDamo14b,BiniDamo14c, KavaOtteWard15, HoppKavaOtte16}.  We have used the 
approach in a series of recent papers \cite{MunnETC20,Munn20,MunnEvan22a,MunnEvan22b}.  We provide, therefore, only 
a brief overview of the method.  

\subsection{Bound orbits and PN compactness parameters}
\label{sec:orbits}

The perturbation is treated as being sourced by a point mass $\mu$ in a bound eccentric geodesic motion 
about a Schwarzschild black hole of mass $M$, with $\mu/M \ll 1$.  We use Schwarzschild coordinates 
$x^{\mu} = (t,r,\theta, \varphi )$ with the line element
\be
ds^2 = -f dt^2 + f^{-1} dr^2
+ r^2 \left( d\theta^2 + \sin^2\theta \, d\varphi^2 \right) ,
\ee
where $f = 1 - 2M/r$.  The coordinates are aligned so that the motion is in the equatorial plane, with four-velocity
\be
\label{eqn:four_velocity}
u^\a(\tau) = \frac{dx_p^{\alpha}(\tau)}{d\tau} 
= \l \frac{{\mathcal{E}}}{f_{p}}, u^r, 0, \frac{{\mathcal{L}}}{r_p^2} \r ,
\ee
where $\mathcal{E}$ and $\mathcal{L}$ are the specific energy and angular momentum, respectively.  We transform from 
parameters $\tau, \mathcal{E},\mathcal{L}$ to Darwin \cite{Darw59} parameters $\chi, p, e$ 
\cite{CutlKennPois94,BaraSago10} via
\begin{align}
\label{eqn:defeandp}
{\mathcal{E}}^2 &= \frac{(p-2)^2-4e^2}{p(p-3-e^2)},
\q
{\mathcal{L}}^2 = \frac{p^2 M^2}{p-3-e^2},  \notag \\
& \qq \q  r_p \l \chi \r = \frac{pM}{1+ e \cos \chi} .
\end{align}
One radial libration corresponds to a $2 \pi$ advance in the relativistic anomaly $\chi$.  The other 
three coordinates (and $\tau$) can be related to $\chi$ via ODEs \cite{HoppEvan10,ForsEvanHopp16}.  The function
$\vp_p(\chi)$ can be expressed analytically in terms of the incomplete elliptic integral of the first kind 
$F(x|m)$ \cite{HoppETC15,GradETC07} and then PN expanded in powers of $1/p$.  In contrast, the integrand for 
$t_p(\chi)$ is first PN expanded and then the result is integrated analytically term by term.  

This representation provides simple means to compute the fundamental frequencies for radial libration, $\O_r$, and 
mean azimuthal motion, $\O_\vp$ (per radial cycle).  Explicitly, the radial period can be derived from
\begin{gather}
\label{eqn:O_r}
T_r = \int_{0}^{2 \pi}  \frac{r_p \l \chi \r^2}{M (p - 2 - 2 e \cos \chi)}
 \left[\frac{(p-2)^2 -4 e^2}{p -6 -2 e \cos \chi} \right]^{1/2}    d \chi ,
 \notag
\end{gather}
with $\Omega_r = 2 \pi/T_r$.  The integrand is readily PN expanded.  The mean azimuthal frequency follows as
\be
\label{eqn:O_phi}
\O_\varphi = \frac{4}{T_r} \left(\frac{p}{p - 6 - 2 e}\right)^{1/2} \, 
K\left(-\frac{4 e}{p - 6 - 2 e}  \right) ,
\ee
where $K(m)$ is the complete elliptic integral of the first kind \cite{GradETC07}, which is also then PN expanded 
in $1/p$.  Finally, the alternative compactness parameter, $y = (M\O_{\vp})^{2/3}$, can be obtained in terms of 
an expansion in $1/p$ and inverted for $p(y)$ as an expansion in $y$.  For eccentric motion, each PN order will itself 
be an added expansion in powers of eccentricity $e$.

\subsection{The RWZ master equations}
\label{sec:TDmasterEq}

In the RWZ formalism \cite{ReggWhee57,Zeri70}, the first-order perturbation of 
Schwarzschild spacetime is encoded by a pair of master equations (one for each parity) that take the 
frequency-domain (FD) form
\be
\label{eqn:masterInhomogFD}
\left[\frac{d^2}{dr_*^2} +\omega^2 -V_l (r) \right]
X_{lmn}(r) = Z_{lmn} (r) .
\ee
Here $r_{*} = r + 2 M \ln | r/2 M - 1 |$ is the tortoise coordinate, $X_{lmn}$ are the mode functions, and the 
frequencies $\o \equiv \o_{mn} = m \O_\vp + n \O_r$ form a discrete spectrum derived from the periodicities in the 
geodesic motion.  The FD source term follows as a Fourier series amplitude
\begin{align}
Z_{lmn} &= \frac{1}{T_r} \int_0^{2 \pi} \Big( G_{lm}(t) \, \delta[r - r_p(t)] \notag \\& \qq \qq \qq
+ F_{lm}(t) \, \delta'[r - r_p(t)] \Big) e^{i \o t} dt .
\end{align}
Several variants of the master equations exist, and we utilize the Zerilli-Moncrief equation for even-parity modes 
and the Cunningham-Price-Moncrief equation for odd-parity modes \cite{MartPois05,HoppEvan10}.  These choices in 
turn give rise to particular forms for $G_{lm}(t)$ and $F_{lm}(t)$.  Due to symmetries in the equatorial source 
motion, for a given $l$ and $m$ only an even-parity or odd-parity mode will exist depending upon whether $l+m$ is 
an even or odd integer, respectively.  The potential $V_{l}(r)$ is also parity dependent, being either the Zerilli 
(even) or Regge-Wheeler (odd) potential.

The homogeneous form of these equations yields two independent solutions: $X_{lmn}^{-} = X_{lmn}^{\rm in}$, with 
causal (downgoing wave) behavior at the horizon, and $X_{lmn}^{+} = X_{lmn}^{\rm up},$ with causal (outgoing wave) 
behavior at infinity.  The odd-parity homogeneous functions can be determined directly using the MST formalism
\cite{ManoSuzuTaka96a}, which we summarize next.  The even-parity counterparts are derived using the trick 
\cite{Munn20,ForsEvanHopp16} of solving the Regge-Wheeler equation for the ``wrong parity'' and then using those 
solutions to derive the even-parity modes through use of the Detweiler-Chandrasekhar transformation 
\cite{Chan75,ChanDetw75,Chan83,Bern07}. 

\begin{widetext}
\subsection{The MST solutions to the homogeneous master equation}

The MST solution \cite{ManoSuzuTaka96a,SasaTago03} for $X_{lmn}^+$ can be expressed as
\begin{align}
\label{eqn:XupMST}
X_{lmn}^+ = e^{iz} z^{\nu+1} \left(1- \frac{\e}{z}\right)^{-i \e} 
\sum_{j=-\infty}^{\infty} a_j^{\nu} (-2 i z)^{j} \frac{ \G(b-2) \G(b) }{\G(b^*+2) \G(b^*) } U(b,c, -2 i z) ,
\end{align}
where $b = j + \nu + 1 - i \e$ and $c = 2j + 2 \nu + 2$ (see also \cite{KavaOtteWard15}).  In this equation, $U$ is 
the irregular confluent hypergeometric function, $\e = 2 M \o \eta^3$, $z = r \o \eta$, with $\eta = 1/c$ being a 
reintroduced (0.5)PN parameter.  To find a solution, $\nu$ (the renormalized angular momentum) and series 
coefficients $a_j$ are ascertained through a continued fraction method \cite{ManoSuzuTaka96a,SasaTago03}, with 
the eigenvalue for $\nu$ allowing the series to converge on both ends.  As previously discussed in earlier 
applications \cite{Munn20,MunnEvan22a,MunnEvan22b}, these parameters and coefficients are (PN) expanded in 
powers of $\e$, and the full solutions have expansions in both $z$ and $\e$.

In a similar fashion the inner, or horizon, solution, $X_{lmn}^-$, is given by
\begin{align}
\label{eqn:XinMST}
X_{lmn}^- = e^{-iz} \left( \frac{\e}{z} \right)^{i \e+1} \left(1- \frac{\e}{z}\right)^{-i \e} 
\sum_{j=-\infty}^{\infty} a_j^{\nu}  \frac{ \Gamma(g) \Gamma(h) }{\Gamma(k) } \, {}_2F_1(g,h, k,1-z/\e) ,
\end{align}
where $g = j + \nu - 1 - i \e, h = -j - \nu - 2 - i\e$, and $k = 1 - 2 i \e$.  The quantities $\nu$ and $a_j$ 
appearing here are the same as those that arise in the outer solution \eqref{eqn:XupMST}.  The process of PN 
expanding both of these homogeneous solutions by collecting on powers of $\eta$ is fully described in 
\cite{Munn20}, based on the methods presented in \cite{KavaOtteWard15}.   

When the RWZ mode functions are computed in this manner, the normalization is typically set by having taken 
$a_0 = 1$ at the start of the recursion calculation.  The resulting amplitudes at infinity and the horizon 
\be
X^{\pm}_{lmn} \sim A^{\pm}_{lmn} \, e^{\pm i \o r_*} ,
\ee
will be such that $|A^{\pm}_{lmn}| \ne 1$.  To simplify (at least the presentation of) the flux calculations, 
it is convenient to adopt unit-normalized modes, with $\hat{X}^{\pm}_{lmn} \sim \exp(\pm i \o r_*)$ 
\cite{HoppEvan10,ForsEvanHopp16}.  The initial amplitudes can be found, respectively, by taking the limit as $z$ in
\eqref{eqn:XupMST} goes to infinity and as $z$ in \eqref{eqn:XinMST} approaches the horizon.  Then, we 
find
\begin{align}
\hat{X}^{\rm +}_{lmn} &= \frac{e^{iz} (-2 i z)^{\nu+1}}{A_{lmn}^{+, \rm sum}} (-2 i \e)^{-i \e} 
\left(1- \frac{\e}{z}\right)^{-i \e}
\sum_{j=-\infty} a_j^{\nu} (-2 i z)^{j} \frac{ \G(b-2) \G(b) }{\G(b^*+2) \G(b^*) }
U(b,c, -2 i z),   \notag \\
A_{lmn}^{+,\rm sum} &= \sum_{j=-\infty} a_n^{\nu} \frac{ \G(j + \nu - 1 - i \e) \G(j + \nu + 1 - i \e) }{\G(j + \nu + 3 + i \e) \G(j + \nu + 1 + i \e) },  \notag \\
\hat{X}^{-}_{lmn} &= X^{-}_{lmn}/A^{-}_{lmn}, \qq \qq A^{-}_{lmn} = \sum_{j=-\infty} a_j^{\nu}  
\frac{ \G(j+\nu-1- i\e) \G(-j - \nu - 2 - i \e) }{\G(1-2 i \e) } .
\end{align}
The renormalized mode functions, $\hat{X}^{\rm +}_{lmn}$ and $\hat{X}^{\rm -}_{lmn}$, can be PN expanded just as 
with the original modes.  However, while it is convenient to think of $\hat{X}^{\rm +}_{lmn}$ and 
$\hat{X}^{\rm -}_{lmn}$ for purposes of introducing the flux calculations, from a symbolic computational standpoint 
it is more efficient to work with specific factorized versions of these functions, as described in the next 
subsection.

\subsection{The horizon fluxes}

Using the unit-normalized homogeneous solutions for the mode functions to construct the Green function, 
integration over the point-particle source yields the following set of normalization coefficients (or asymptotic 
amplitudes)
\begin{align}
\label{eqn:EHSC}	
C_{lmn}^\pm  = \frac{1}{W_{lmn} T_r} \int_0^{T_r}  \bigg( \frac{dt}{d \chi} \bigg)
\Bigg[ \frac{1}{f_{p}} \hat X^\mp_{lmn}
 G_{lm} + \l \frac{2M}{r_{p}^2 f_{p}^{2}} \hat X^\mp_{lmn}
 - \frac{1}{f_{p}} 
 \frac{d \hat X^\mp_{lmn}}{dr} \r F_{lm}
 \Bigg]  e^{i \o t}  \, d\chi .
\end{align}
Here, $W_{lmn}$ is the Wronskian
\be
W_{lmn} = f \hat{X}^-_{lmn} \frac{d \hat{X}^+_{lmn}}{dr}
- f \hat{X}^+_{lmn} \frac{d \hat{X}^-_{lmn}}{dr} .
\ee
Once the $C_{lmn}^\pm$ amplitudes are computed, the time domain solutions and fluxes can be obtained.  In principle, 
to expand \eqref{eqn:EHSC} analytically, the homogeneous solutions are evaluated at the location of the particle using 
the PN-expanded and $e$-expanded geodesic motion of the secondary.  Then, expansions in $y$ (or $1/p$) and $e$ are 
generated for the remaining parts of the integrand.  Integration term by term produces a double expansion for each 
$C_{lmn}^\pm$.  See \cite{Munn20,HoppKavaOtte16} for a more details.

Our concern in this paper is with the horizon-side coefficients, $C^{-}_{lmn}$, which can be used to obtain the 
rate at which energy and angular momentum are absorbed by the black hole according to
\begin{align}
\label{eqn:BHPThorfluxes}
\left\langle \frac{dE}{dt} \right\rangle_{\rm H} = 
\sum_{lmn}\frac{\o^2}{64\pi}\frac{(l+2)!}{(l-2)!} |C^{-}_{lmn}|^2  \qquad\qquad
\left\langle \frac{dL}{dt} \right\rangle_{\rm H} = 
\sum_{lmn}\frac{m \o}{64\pi}\frac{(l+2)!}{(l-2)!} |C^{-}_{lmn}|^2 .
\end{align} 

However, as discussed in \cite{Munn20}, the straightforward implementation of this procedure produces symbolic 
expressions of unwieldy size, limiting the attainable PN order and order in the eccentricity expansion.  The 
computational task is reduced drastically by removing certain $z$-independent factors from the homogeneous 
solutions prior to calculating the source integrals, and then multiplying those factors back in at the end.  The 
factors for $X^+_{lmn}$, relevant to computing the flux at infinity, are given in \cite{Munn20} and exactly match
the $S_{lmn}$ tail factorization (N.~Johnson-McDaniel, private communication) that generalized for eccentric 
orbits the circular-orbit $S_{lm}$ factorization explored by Johnson-McDaniel in \cite{JohnMcDa14}.  Not 
surprisingly, a similar factorization exists on the horizon side.  Removing the $z$-independent factors from 
$X^-_{lmn}$, which affects the Wronskian, modifies \eqref{eqn:EHSC} and leads to an altered set of coefficients 
$\tilde{C}_{lmn}^{-}$.  The factor pulled out is seen in the relationship
\begin{align}
C_{lmn}^{-} = \left(\frac{2}{p} \right)^{\D \nu} \frac{\G(1 + \D \nu - i \e)^2}{\G(1 + 2 \D\nu) \G(1 - 2 i \e)} 
\tilde{C}_{lmn}^{-} .
\end{align}
Here, $\D\nu = \nu - l$.  The fluxes, for each mode, are then recovered by multiplying back in the complex square 
of this factor
\begin{align}
| C_{lmn}^{-} |^2 &=  \left(\frac{2}{p} \right)^{2 \D \nu} \frac{\G(1 + \D \nu - i \e)^2 \G(1 + \D \nu + 
i \e)^2}{\G(1 + 2 \D\nu)^2 \G(1 - 2 i \e) \G(1 + 2 i \e)}  | \tilde{C}_{lmn}^{-} |^2 .
\end{align}
Note that in the PN limit, complex values of $\nu$ are never encountered.  The use of this factorization 
significantly improves our ability to reach high PN order.  A few computational benchmarks using this 
procedure are given in Table~\ref{tab:Clmntimes}.

\begin{table*}[t]
\begin{center}
\caption{Overview of the computational time needed for expansion of various even-parity normalization constants to 
high PN order.  Expansions were found for specific $l$ but general $m$ and $n$ on the UNC Longleaf cluster.  The 
third and fourth columns indicate the time and memory, respectively, needed for the calculation.  The fifth column 
gives the approximate size of a text file holding the output.  In each case the comparable odd-parity computation
is simpler and faster. }
\vspace{1em}
\label{tab:Clmntimes}
\begin{tabular}{ || c | c | c | c | c ||}
\hline\hline
Coefficient & Relative Order & CPU time (hours) & Memory & Text File Size \\
\hline
$C^{-}_{2mn}$ &  18PN/$e^{10}$  &  81.7   &    5GB  &  140MB \\
\hline
$C^{-}_{4mn}$ &  14PN/$e^{10}$   &   21.1   &   3GB  &   50MB  \\
\hline
$C^{-}_{2mn}$ & 10PN/$e^{20}$   &   3.3    &   2GB   &  120MB  \\
\hline \hline
\end{tabular}
\end{center}
\end{table*}

\section{PN expansion of the horizon energy absorption to 18PN}
\label{sec:EfluxExps}

As mentioned in Sec.~\ref{sec:intro}, recent work by Isoyama et al.~\cite{IsoyETC22} found and utilized the horizon 
flux for eccentric-orbit EMRIs to 5PN, at $e^{10}$, relative to the leading flux at infinity.  Recall that for a Kerr 
primary, this result is 2.5PN relative to the leading horizon flux.  For a Schwarzschild primary, it is only 1PN 
relative to the dominant horizon flux.  Less well known is that Forseth, in his thesis \cite{Fors16} (see sections 
7.3 and 7.4; also see the posted APS talk \cite{Fors16b} and Capra talk \cite{Evan16}), found the exact-in-$e$ 
horizon absorption for nonspinning 
EMRIs (at lowest order in the mass ratio) through 2PN relative to the leading horizon flux (which we will henceforth 
refer to as \emph{horizon-relative}) and a couple additional exact-in-$e$ terms and accurate numerical coefficients to 
high order in $e$ up to 7PN horizon-relative.  In other words, Forseth's work already had mixed analytic/numerical 
understanding of the eccentric-orbit horizon flux to 11PN.

To begin enumeration of our findings, it is useful to recall that high-order work on circular-orbit horizon fluxes 
\cite{Shah14, Fuji15} determines the expected form of the PN expansion.  However, individual coefficients at each 
PN order are now replaced by functions of $e$ in the eccentric-orbit case.  The leading part of the horizon energy 
flux is given by the expansion
\begin{align}
\label{eqn:energyfluxHor}
\left\langle \frac{dE}{dt} \right\rangle_H =  \frac{32}{5} \left(\frac{m_2}{m_1}\right)^2 y^9 &
\biggl[\mathcal{B}_0 + y\mathcal{B}_1 + y^2\mathcal{B}_2
+ y^3\biggl(\mathcal{B}_3+\mathcal{B}_{3L}\log y \biggr)
+ y^4\biggl(\mathcal{B}_4+\mathcal{B}_{4L}\log y \biggr)
\notag\\&\quad+ y^5\biggl(\mathcal{B}_5+\mathcal{B}_{5L}\log y \biggr)
+ y^{11/2}\mathcal{B}_{11/2} + y^6\biggl(\mathcal{B}_6
+\mathcal{B}_{6L}\log y +\mathcal{B}_{6L2}\log^2 y \biggr)
\notag\\&\quad+ y^{13/2}\mathcal{B}_{13/2}
 + y^7\biggl(\mathcal{B}_7 +\mathcal{B}_{7L}\log y +\mathcal{B}_{7L2}\log^2 y \biggr)+\cdots \biggr] , 
\end{align}
where the $\mathcal{B}_k$, $\mathcal{B}_{kL}$, $\mathcal{B}_{kL2}$, etc, are functions of $e$.  The structure of 
this expansion differs from that of the flux at infinity, principally with the half-integer (tail) term not showing 
up until 5.5PN horizon-relative, which is at 9.5PN in the total flux.  This contrasts with the infinity-side flux 
where the tail appears at 1.5PN.  

In discussing the horizon fluxes, we find it convenient to refer to terms by their horizon-relative order.  
Thus, in our labeling of the eccentricity enhancement functions, $\mathcal{B}_k$, $\mathcal{B}_{kL}$, etc, the 
integer or half-integer $k$ reads out directly the horizon-relative order.  From this point on we will implicitly 
refer to PN terms by their horizon-relative order.  The structure beyond what is displayed in 
\eqref{eqn:energyfluxHor} is clear; at integer orders, a new power of $\log y$ shows up every 3PN in the expansion 
and the first $\log y$ term at half-integer order will appear at 8.5PN, with added powers of log at 11.5PN, 
14.5PN, etc.

In 2016, Forseth \cite{Fors16} used the numeric-analytic fitting procedure discussed in \cite{ForsEvanHopp16} to 
fit for coefficients in power series expansions in $e^2$ of a number of these enhancement functions.  The accurate 
numerical results reached 7PN order.  For energy fluxes, he found closed-form expressions for $\mathcal{B}_0(e)$, 
$\mathcal{B}_1(e)$, $\mathcal{B}_2(e)$, $\mathcal{B}_{3L}(e)$, and $\mathcal{B}_{4L}(e)$.  He then extracted 
analytic coefficients for terms in truncated power series expansions in $e^2$ for many of the remaining terms to 
7PN, specifically computing $\mathcal{B}_3(e)$ to $e^{40}$, $\mathcal{B}_4(e)$ to $e^{4}$, $\mathcal{B}_{5L}(e)$ 
to $e^{30}$, $\mathcal{B}_{11/2}(e)$ to $e^{12}$, $\mathcal{B}_{6L2}(e)$ to $e^{12}$, $\mathcal{B}_{13/2}(e)$ to 
$e^{4}$, and $\mathcal{B}_{7L2}(e)$ to $e^{6}$.  No new analytic coefficients were found in $\mathcal{B}_5(e)$, 
$\mathcal{B}_6(e)$, $\mathcal{B}_{6L}(e)$, $\mathcal{B}_{7}(e)$, or $\mathcal{B}_{7L}(e)$ beyond the already 
known circular-orbit limit, but additional accurate numeric coefficients were found. 

This paper now extends the work in \cite{Fors16} using the analytic expansion methods of \cite{KavaOtteWard15,
HoppKavaOtte16, Munn20}.  The result is a pair of expansions, with one derived to $e^{20}$ through 10PN and 
the other to $e^{10}$ through 18PN.  (These results require doing extensive symbolic computations with 
\textsc{Mathematica} on a cluster computer and we found it useful to press the expansions as deeply as possible 
alternately in PN order and in powers of $e$.)  Again, because the horizon flux is suppressed by 4PN, our 
calculations go to 14PN and 22PN, respectively, relative to the leading $\mathcal{L}_0(e)$ and $\mathcal{J}_0(e)$ 
quadrupole fluxes \cite{MunnETC20}.  

We now step through a presentation of each energy flux eccentricity enhancement function through 8PN order.  
Each function was computed through $e^{20}$, but some were found to resum into closed-form expressions and several 
are truncated here to fewer terms than $e^{20}$ for brevity.  The full results through 10PN/$e^{20}$ and 
18PN/$e^{10}$ are posted online \cite{BHPTK18,UNCGrav22}.  As we mentioned, a subset of 
these coefficients were presented in \cite{Fors16}.  At the lowest few orders, closed-form expressions were found 
through 2PN \cite{Fors16}
\begin{align}
\mathcal{B}_0 = &\frac{1}{(1-e^2)^{15/2}}
\left(1+\frac{31}{2}e^2+\frac{255}{8}e^4+\frac{185}{16}e^6
+\frac{25}{64}e^8\right),
\\\mathcal{B}_1 = &\frac{1}{(1-e^2)^{17/2}}
\left(4+\frac{147}{2}e^2+\frac{799}{8}e^4-\frac{2635}{16}e^6
-\frac{13515}{128}e^8-\frac{275}{64}e^{10}\right),\\
\mathcal{B}_2 = &\frac{1}{\left(1-e^2\right)^{19/2}}
\biggl(-\frac{181}{14}+\frac{1336}{21}e^2
+\frac{25097}{24}e^4+\frac{42743}{48}e^6+\frac{489245}{768}e^8  \notag \\&\qq
+\frac{360197}{768}e^{10} +\frac{6025}{256}e^{12}\biggr)  +\frac{75}{2 \sqrt{1-e^2}} \mathcal{B}_0.
\end{align}
Interestingly, the $\mathcal{B}_0(e)$ enhancement function has been recently separately uncovered by Datta 
\cite{Datt23}.  The 2PN function is given here in a form that is slightly different from what was shown in 
\cite{Fors16}.  As with the 2PN flux at infinity \cite{MunnETC20}, the polynomial attached to the subdominant 
singular factor can be manipulated into a term proportional to a lower-order function, in this case $\mathcal{B}_0(e)$.

At 3PN we found that the non-log series, already known to $e^{40}$ \cite{Fors16}, could be put into a closed form.  
Among several steps in this reduction, we first notice that the 3PN log term reappears in the 3PN non-log function.  
The overall complicated closed-form expression at 3PN is reminiscent of the 3PN flux at infinity \cite{MunnEvan19a}, 
though the present one lacks a $\chi(e)$-like \cite{ArunETC08a} (infinite series) function.  The 3PN non-log and 
3PN log functions are 
\begin{align}
\label{eqn:B3}
\mathcal{B}_3 = &\frac{1}{\left(1-e^2\right)^{21/2}}\biggl[-\frac{9530309}{51975}-\frac{14041757 e^2}{4725}
-\frac{81025787 e^4}{8400}-\frac{102162779 e^6}{3600}-\frac{188105821 e^8}{5760} -\frac{4984577 e^{10}}{600} 
\notag \\&    -\frac{2917799 e^{12}}{2048}-\frac{82525 e^{14}}{1024} +
\sqrt{1-e^2} \bigg(\frac{4348832}{14553}+\frac{4081074097 e^2}{727650}+\frac{29035617361 e^4}{2910600}
-\frac{6084796133 e^6}{2328480}   \notag \\&
+\frac{115633347503 e^8}{18627840}+\frac{230334470711 e^{10}}{46569600}
+\frac{12625 e^{12}}{64} \bigg) \biggr]  -
\left[\frac{35}{107}\pi^2 + \log\left(\frac{1-e^2}{1+\sqrt{1-e^2}}\right)\right]\mathcal{B}_{3L},\\
\label{eqn:B3L}
\mathcal{B}_{3L} =& \frac{-1}{(1-e^2)^{21/2}}\biggl(
\frac{1712}{105}+\frac{79822}{105} e^2+\frac{393867}{70}e^4
+\frac{110103}{10}e^6+\frac{100687}{16}e^8+\frac{287937}{320} e^{10}
+\frac{3745}{256} e^{12}
\biggr) .
\end{align}

The 4PN non-log term marks the first appearance of additional transcendental numbers, such as $\gamma_E$ and 
$\log 3$.  This series has no overall closed form.  However, there are parts within it proportional to $\pi^2$ and 
$\g_E$ that do terminate in finite polynomials, which can be seen in the following expansion truncated at 
$e^{16}$
\begin{align}
\label{eqn:B4}
\mathcal{B}_{4} =& \frac{1}{(1-e^2)^{23/2}}\biggl[ \frac{10859497}{22050} + \frac{52}{3}\pi^2
-\frac{1024}{15}\gamma_E - \frac{3980}{21} \log(2) + \bigg(\frac{6799223}{420}+ 420 \pi ^2
-\frac{14992}{3}\gamma_E  \notag\\& 
-\frac{ 77324}{15} \log(2)-\frac{30618}{5}\log(3)\bigg)e^2 +\bigg(\frac{16251413749}{176400}
-\frac{4045}{3}\pi^2-\frac{293664}{5}\gamma_E -\frac{12610811}{35} \log(2)  \notag\\&
+98415 \log(3)\bigg)e^4  + \bigg( \frac{8038555537}{16800}-\frac{975574 \g_E}{5}-\frac{100681 \pi ^2}{4}
+\frac{2565846047 \log(2)}{540}-\frac{118403451 \log(3)}{160}  \notag \\&
-\frac{1650390625 \log(5)}{864} \bigg) e^6 + \bigg(  \frac{122640123477}{89600}-\frac{1083614 \g_E}{5}
-\frac{1767591 \pi ^2}{32} - \frac{77581480669 \log(2)}{1440}    \notag \\&
-\frac{8528452173  \log(3)}{640}+\frac{36513671875 \log(5)}{1152} \bigg) e^8  
+ \bigg( \frac{112982456593}{89600}-\frac{3195171 \g_E}{40}-32732 \pi^2  \notag \\&
+\frac{12472278232151 \log(2)}{27000} +\frac{280561035495519 \log(3)}{1024000} 
-\frac{18301802734375 \log(5)}{73728}    \notag \\&
-\frac{3160616408486747 \log(7)}{27648000}  \bigg) e^{10} + \bigg( \frac{3835468188137}{6451200}
-\frac{39893 \g_E}{5} -\frac{152859 \pi^2}{32}   \notag \\&
-\frac{4696393122910219 \log(2)}{972000}   
-\frac{4645413320517747 \log(3)}{2048000} 
+\frac{4858337978515625 \log(5)}{3981312}     \notag \\&
+\frac{330478628708225893 \log(7)}{165888000}  \bigg) e^{12}  
+ \bigg( \frac{1532279204891}{4515840}  +\frac{12163662267614069749 \log(2)}{254016000}  
-\frac{6125 \g_E}{64}  \notag\\&
+\frac{8312610758654209851 \log(3)}{1605632000} -\frac{6592628145458984375 \log(5)}{1560674304}   
-\frac{262549480209071768377 \log(7)}{15925248000}    \notag \\&
-\frac{40425 \pi^2}{512}  \bigg) e^{14}  
+ \bigg( \frac{19581861707467}{82575360}-\frac{844848484081061147 \log(2)}{2381400}  
+\frac{195002386788591269793 \log(3)}{2569011200}  \notag \\&
-\frac{13605756642412109375 \log(5)}{4161798144}
+\frac{439388545040978324393 \log(7)}{5096079360}  \bigg) e^{16}  + \cdots  \biggr] .
\end{align}
The 4PN log term has the previously known closed form
\begin{align}
\label{eqn:B4L}
\mathcal{B}_{4L} =& \frac{1}{(1-e^2)^{23/2}}\biggl(-\frac{9148}{105}
-\frac{11348}{3}e^2-\frac{2650657}{105}e^4-\frac{412167 e}{20}e^6
+\frac{9681067}{160}e^8+\frac{4810141}{80}e^{10}  \notag\\& \q
+\frac{1698271}{160}e^{12} +\frac{99085}{512} e^{14}\biggr) .
\end{align}

At 5PN, the non-log function is similar in form to its 4PN counterpart, though without any apparent truncation in 
the parts of the series attached to $\pi^2$ and $\g_E$.  As a result, we only present the first few powers of 
$e^2$ here.  We find that the 5PN log term can be put into a closed-form expression, with a subdominant singular 
term that involves the recurrence of the 3PN log function
\begin{align}
\label{eqn:B5}
\mathcal{B}_5 =& \frac{1}{(1-e^2)^{25/2}}\biggl[ \frac{2547493}{1372}
+\frac{2528}{35}\gamma_E+\frac{780}{7}\pi^2-\frac{425612}{2205}\log(2)  
+ \bigg(  \frac{1328278289}{61740}+\frac{1281872 \g_E}{45}+\frac{359584 \pi ^2}{63}  \notag \\&
+\frac{8972492 \log(2)}{735}+\frac{137781 \log(3)}{5} \bigg) e^2  
+ \bigg( -\frac{1304314425913}{1234800}+\frac{210446456 \g_E}{315}+\frac{12945643 \pi ^2}{252}  \notag \\&
+\frac{22816656827 \log(2)}{8820}-\frac{7158051 \log(3)}{20}  \bigg) e^4   
+ \bigg( -\frac{1963383791933}{231525}+\frac{61546622 \g_E}{15}+\frac{9407723 \pi ^2}{56}  \notag \\&
-\frac{4622401802581 \log(2)}{158760}+\frac{4851838359 \log(3)}{2240}
+\frac{660796484375 \log(5)}{36288} \bigg) e^6  + \bigg(  -\frac{91961130977371}{3763200}   \notag \\&
+\frac{936331714 \g_E}{105}+\frac{38160541 \pi ^2}{96}
+\frac{13904921619359 \log(2)}{30240}+\frac{349567982667 \log(3)}{1792}  \notag \\&
-\frac{15006157421875 \log(5)}{48384} \bigg) e^8  +
\cdots\biggr],\\
\label{eqn:B5L}
\mathcal{B}_{5L} =& \frac{1}{(1-e^2)^{25/2}}\biggl(\frac{27212}{49}+\frac{15715351 e^2}{441}
+\frac{1829922349 e^4}{4410}+\frac{9960979829 e^6}{5880}+\frac{9496143917 e^8}{3360} \notag \\&
+\frac{3196758989 e^{10}}{1920}+\frac{952203067 e^{12}}{5376}-\frac{205586393 e^{14}}{7168}
-\frac{4047085 e^{16}}{4096} \biggr) + \frac{105}{2 \sqrt{1-e^2}} \mathcal{B}_{3L}  .
\end{align}

As already known from the circular-orbit limit, 5.5PN marks the first appearance of a half-integer flux term.  
As expected from our experience with the infinity-side flux, the 5.5PN term appears to be an infinite series with 
rational number coefficients (once an overall factor of $\pi$ is pulled out)
\begin{align}
\mathcal{B}_{11/2} =& \frac{\pi}{(1-e^2)^{13}}\biggl(-\frac{109568}{1575}
-\frac{72974}{9}e^2-\frac{11159458}{75}e^4-\frac{2064646855}{2592}e^6
-\frac{556894606109}{362880} e^8  \notag\\&
-\frac{1634615689436141}{1451520000}e^{10}-\frac{647533375166093}{2177280000}e^{12} 
-\frac{614373168703875323 e^{14}}{27311800320000} \notag\\& 
-\frac{139393544295440923 e^{16}}{655483207680000}
+\frac{6810696714424201 e^{18}}{3398024948613120000}
+\frac{261931344108584947 e^{20}}{84950623715328000000} + \cdots \biggr) .
\end{align}
The specific eccentricity singular factor (power of $1 - e^2$) ensures that the power series converges as 
$e \rightarrow 1$.

At 6PN, in the non-log term, there is a significant increase in coefficient complexity.  Only a few coefficients are
listed here for brevity, with the rest available elsewhere \cite{BHPTK18,UNCGrav22}.  Experience leads us to believe 
that the 6PN log term will likely yield a closed form, but the expansions would have to be computed beyond $e^{20}$ to 
confirm the conjecture and to solve for the (presumed) rational number polynomial.  A key part of that belief is 
that we are able to simplify the appearance of the full term by isolating the transcendental numbers.  The 6PN 
$\log^2$ term, meanwhile, is an additional closed-form function, which reappears in the 6PN log term
\begin{align}
\mathcal{B}_{6} =& \frac{1}{(1-e^2)^{27/2}}\biggl[ \frac{278408801583211}{48134047500}
-\frac{594376 \g_E}{2835}+\frac{17112988 \pi ^2}{33075}-\frac{256 \pi ^4}{45}
-\frac{1485396268 \log(2)}{694575}    \notag \\&
-\frac{27392}{315} \pi ^2 \log(2)+\frac{1465472 \log^2(2)}{11025} -\frac{54784 \zeta (3)}{105} 
+ \bigg( -\frac{3184621738776971}{96268095000}-\frac{209233244 \g_E}{2835}  \notag \\&
+\frac{488155076 \pi ^2}{33075}-\frac{30992 \pi ^4}{45}-\frac{29196601796 \log(2)}{231525}
-\frac{3316144}{315} \pi ^2 \log(2)+\frac{177413704 \log^2(2)}{11025}  \notag \\&
-\frac{592677 \log(3)}{7} - \frac{6632288 \zeta (3)}{105}  \bigg) e^2 + \bigg(-\frac{3155544818212373}{729303750} 
-\frac{534255044 \g_E}{189}-\frac{7719987437 \pi ^2}{44100}     \notag \\&
-\frac{195424 \pi ^4}{15} -\frac{30309834111577 \log(2)}{2778300}-\frac{20910368}{105} \pi ^2 \log(2)   
+\frac{1118704688 \log^2(2)}{3675}+\frac{385509051 \log(3)}{280}    \notag \\&
-\frac{41820736 \zeta (3)}{35} \bigg) e^4
+ \bigg( \frac{2981094178946939}{466754400}  
- \frac{9377211302 \g_E}{315} -\frac{28572566131 \pi ^2}{10584}-71670 \pi ^4   \notag \\&
+\frac{216371435533139  \log(2)}{2000376}  
-\frac{7668690}{7} \pi ^2 \log(2)+\frac{82054983 \log^2(2)}{49}-\frac{24002939547 \log(3)}{2240}   \notag \\&
-\frac{27901272734375 \log(5)}{326592}-\frac{46012140 \zeta (3)}{7}  \bigg) e^6
+\cdots \biggr],\\
\mathcal{B}_{6L} =& \frac{1}{(1-e^2)^{27/2}}\biggl(-\frac{1265945848}{694575}-\frac{69768064262 e^2}{694575}
-\frac{1168647880507 e^4}{926100}-\frac{16908108709883 e^6}{1852200}   \notag \\&
-\frac{286763142507587 e^8}{7408800}
-\frac{69524389377859 e^{10}}{1008000}-\frac{206416658766499 e^{12}}{4704000}  
-\frac{417391782384293 e^{14}}{56448000}   \notag \\&
+\frac{40129383119743 e^{16}}{52684800}   
+\frac{14409668299744981 e^{18}}{22759833600}+\frac{192650436984624487 e^{20}}{455196672000}
+\cdots \biggr)   \notag \\&
-2 \left[\frac{35}{107}\pi^2  + \log\left(\frac{1-e^2}{1+\sqrt{1-e^2}}\right)\right] \mathcal{B}_{6L2}  ,  \\
\mathcal{B}_{6L2} =& \frac{1}{(1-e^2)^{27/2}}\biggl(\frac{1465472}{11025}+\frac{177413704 }{11025}e^2
+\frac{1118704688}{3675} e^4 + \frac{82054983}{49} e^6+\frac{9729806711}{2940} e^8  \notag\\& \q
+\frac{995250121 }{400}e^{10} +\frac{5650230337 }{8400}e^{12} 
+\frac{24483033907 e^{14}}{470400}+\frac{515205 e^{16}}{1024} \biggr).
\end{align}

Like the function at 5.5PN order, the 6.5PN enhancement function is an apparently infinite series with rational 
number coefficients 
\begin{align}
\mathcal{B}_{13/2} =& \frac{\pi}{(1-e^2)^{14}}\biggl( \frac{7239376}{33075}+\frac{2576822347 e^2}{33075}
+\frac{327802444819 e^4}{132300}+\frac{58650165206431 e^6}{2721600}     \notag \\&
+\frac{5301531761061667 e^8}{76204800}   
+\frac{5675365062629170939 e^{10}}{60963840000}
+\frac{38144747001238113839 e^{12}}{731566080000}    \notag \\&
+\frac{2578526099214361612987 e^{14}}{229419122688000}   
+\frac{80254088065124092712893 e^{16}}{110121178890240000}   \notag \\&
+\frac{175336474661571444475081 e^{18}}{28543409568350208000}   
-\frac{99566880013911101100559  e^{20}}{14271704784175104000000} +\cdots \biggr) .
\end{align}

At 7PN we see echoes of lower-order behavior.  First, the 7PN $\log^2$ term is found to have a closed form.  Then 
the 7PN log term is found to have a structure similar to $\mathcal{B}_4(e)$.  Finally, the 7PN non-log function 
displays another increase in complexity, which requires us to truncate its presentation here to just the first few 
coefficients
\begin{align}
\mathcal{B}_{7} =&  \frac{1}{(1-e^2)^{29/2}} \biggl[
\frac{139830180452857}{4202178750}-\frac{41122541072 \g_E}{5457375}+\frac{876544 \g_E^2}{1575}
+\frac{1921663 \pi ^2}{945}-\frac{16384 \g_E \pi ^2}{45}-\frac{784 \pi ^4}{45}  \notag \\&
-\frac{121809908713 \log(2)}{5457375}+\frac{1753088}{525} \g_E\log(2)
-\frac{44752}{45} \pi ^2 \log(2) +\frac{7653496 \log^2(2)}{1575}-\frac{1458 \log(3)}{49} - \notag \\&
\frac{40352 \zeta (3)}{15}   
+ \bigg( \frac{3256393785622158277}{1058949045000}-\frac{21584177896 \g_E}{18375}
+\frac{22831232 \g_E^2}{225}+\frac{5940230269 \pi ^2}{66150}  +\frac{4544 \pi ^4}{15}  \notag \\&
-\frac{2987264 \g_E \pi ^2}{45}  -\frac{367416}{5} \pi ^2 \log(3)
+\frac{121017856}{315} \g_E \log(2)
-\frac{17215808}{315} \pi ^2 \log(2)+\frac{59553632 \log^2(2)}{245}   \notag \\&
-\frac{103840639180343 \log(2)}{76403250}
-\frac{185822334009 \log(3)}{134750}
+\frac{5616216}{25} \g_E \log(3)+\frac{11232432}{25} \log(2) \log(3)  
 \notag \\&
+\frac{2808108 \log^2(3)}{25} -\frac{3598720 \zeta (3)}{21}   \bigg) e^2
+\bigg( \frac{684888927873400705183}{8471592360000}-\frac{6024298579594 \g_E}{218295}  \notag \\&
+\frac{58826032 \g_E^2}{21}   
+\frac{229149574331 \pi ^2}{132300}-\frac{5497760 \g_E \pi ^2}{3} +\frac{2064232 \pi ^4}{45}
-\frac{7386902817169663 \log(2)}{50935500}   \notag \\&
+\frac{6925498816}{225} \g_E \log(2)   
-\frac{263705832}{35} \pi ^2 \log(2)+\frac{145436677388 \log^2(2)}{2205}
+\frac{61967753165457 \log(3)}{2156000}   \notag \\&
-\frac{749764836}{175} \g_E \log(3)  
+\frac{7007148}{5} \pi ^2 \log(3)-\frac{1499529672}{175} \log(2) \log(3)  
-\frac{374882418 \log^2(3)}{175}   \notag \\&
-\frac{6201171875 \log(5)}{42336}   -\frac{135519152 \zeta (3)}{105}  \bigg) e^4  +\cdots \biggl] , \\
\mathcal{B}_{7L} =&  \frac{1}{(1-e^2)^{29/2}} \biggl[ -\frac{4017866767}{363825}+\frac{876544 \g_E}{525}
-\frac{20176 \pi ^2}{45}+\frac{1059728 \log(2)}{225}  + \bigg( -\frac{14891078894803}{15280650}   \notag \\&
+\frac{22831232 \g_E}{75}-\frac{1799360 \pi ^2}{63} + \frac{3959903936 \log(2)}{11025}
+\frac{8424324 \log(3)}{25}  \bigg) e^2  + \bigg( -\frac{650869974249181}{30561300}   \notag \\&
+\frac{58826032 \g_E}{7} - \frac{67759576 \pi ^2}{315}+\frac{423623437208 \log(2)}{11025}
-\frac{1124647254 \log(3)}{175}  \bigg) e^4  \notag \\&
+ \bigg( -\frac{55802998904113}{308700}   
+\frac{11977322344 \g_E}{175}  +\frac{49439206 \pi ^2}{35} -\frac{16445422089422 \log(2)}{33075}  \notag \\&
+\frac{162344445777 \log(3)}{2800}    
+\frac{882958984375 \log(5)}{3024}  \bigg) e^6 + \cdots \biggr] , \\
\mathcal{B}_{7L2} =&  \frac{1}{(1-e^2)^{29/2}}\biggl( \frac{1736824}{1575}+\frac{1320376576 e^2}{11025}
+\frac{26787887416 e^4}{11025}+\frac{18315316581 e^6}{1225}   
+\frac{12160538337 e^8}{392}  \notag \\&
+\frac{1024840623571 e^{10}}{58800}-\frac{5553745441 e^{12}}{1200} - \frac{1982418157607 e^{14}}{470400} 
 -\frac{7256809345951 e^{16}}{15052800} -\frac{23588685 e^{18}}{4096} \biggr) .
\end{align}

The flux function at 7.5PN order is another infinite series with rational number coefficients, similar to those 
at 5.5PN and 6.5PN orders
\begin{align}
\mathcal{B}_{15/2} =& \frac{\pi}{(1-e^2)^{15}}\biggl( -\frac{284700044}{297675}-\frac{748979453 e^2}{1575}
-\frac{2529616180321 e^4}{113400}-\frac{279099734426153 e^6}{979776}  \notag \\&
-\frac{2544038708577181267 e^8}{1828915200}  
-\frac{327745583650604808497 e^{10}}{109734912000}
-\frac{1110213896372403035881 e^{12}}{376233984000}  \notag \\&
-\frac{13562049636351151342862933 e^{14}}{10323860520960000} 
-\frac{958544888235761993942724041 e^{16}}{3964362440048640000}   \notag \\&
-\frac{22610106285369446938573477111  e^{18}}{1284453430575759360000}
-\frac{69643516119521541537867215203 e^{20}}{28543409568350208000000}
+ \cdots \biggr) .
\end{align}

Finally, the 8PN flux terms are similar in complexity to their 7PN counterparts.  We first consider the 8PN 
$\log^2$ term.  By comparison with the infinity-side flux \cite{MunnETC20}, this term would be expected to have a 
closed-form expression, with polynomials multiplying dominant and subdominant eccentricity singular factors.  
This conjecture is supported by finding a closed-form expression of exactly this type for the 8PN $\log^2$ angular 
momentum absorption, $\mathcal{D}_{8L2}$ (see Eq.~\eqref{eqn:D8L2}).  Doing the same for energy absorption is 
unfortunately just out of reach since our symbolic computation stopped at $e^{20}$.  To the depth we calculated, 
the series is
\begin{align}
\label{eqn:B8L2}
\mathcal{B}_{8L2} =&  \frac{1}{(1-e^2)^{31/2}} \biggl[ \frac{56356816}{25725}-\frac{8698826708 e^2}{15435}
-\frac{2649929234297 e^4}{77175}-\frac{34990723365323 e^6}{77175}   \notag \\&
-\frac{440076691742413 e^8}{205800} -\frac{432807788333483 e^{10}}{102900}
-\frac{42100421744609 e^{12}}{11760} -\frac{4121679604910071 e^{14}}{3292800}   \notag \\&
-\frac{3758717869097377 e^{16}}{26342400} 
+\frac{495419426564971 e^{18}}{90316800}+\frac{42850078990521 e^{20}}{8028160}
+ \cdots \biggr]  .
\end{align}
Next, the 8PN log term features numerous transcendental numbers, which even a truncated display of the function 
to $e^4$ reveals.  Finally, the 8PN non-log function displays another increase in complexity, with new terms with 
products of transcendental numbers
\begin{align}
\mathcal{B}_{8} =&  \frac{1}{(1-e^2)^{31/2}} \biggl[
\frac{57784184943753058541}{626368360117500}+\frac{1439910584804 \g_E }{496621125}
+\frac{711232 \g_E ^2}{1225}+\frac{2185682224  \pi ^2}{231525}+\frac{120448 \g_E  \pi ^2}{315}  \notag \\&
-\frac{36992 \pi ^4}{315}-\frac{239862025427236 \log(2)}{10429043625}+\frac{7649024 \g_E  \log(2)}{6615}
-\frac{6515456 \pi ^2 \log(2)}{6615}+\frac{24280192 \log^2(2)}{9261}  +   \notag \\&
\frac{1458 \log(3)}{49}  -\frac{6868864 \zeta (3)}{735}  
+ \bigg(  -\frac{3813515315126610126287}{375821016070500}
+\frac{2350080736803092 \g_E }{297972675} 
-\frac{308587184 \g_E ^2}{735}    \notag \\&
+\frac{78753189211 \pi ^2}{277830}+\frac{91013152 \g_E  \pi ^2}{189}
-\frac{11744272 \pi ^4}{945}+\frac{111453562397431 \log(2)}{12835746} 
-\frac{78444528832 \g_E  \log(2)}{33075}    \notag \\&
+\frac{663243088 \pi ^2 \log(2)}{2205}
-\frac{397392460904 \log^2(2)}{231525}+\frac{5146706320929 \log(3)}{700700} 
-\frac{19499292}{25} \g_E  \log(3)    \notag \\&
+\frac{2388204}{5} \pi ^2 \log(3)
-\frac{56004696}{25} \log(2) \log(3)
-\frac{9749646 \log^2(3)}{25}+\frac{148340096 \zeta (3)}{441} \bigg) e^2  \notag \\&
+ \bigg( -\frac{12281508178808572470642587}{15032840642820000}
+\frac{568110967198223434 \g_E }{1489863375}-\frac{261357876236 \g_E ^2}{11025} 
-\frac{323422432 \pi ^4}{945} \notag \\&
-\frac{70667930236759 \pi ^2}{5556600}+\frac{4468412344 \g_E  \pi ^2}{189}
+\frac{3702433771498464049 \log(2)}{2528253000} 
-\frac{7848597389776 \g_E  \log(2)}{33075}   \notag \\&
+\frac{543204481448 \pi ^2 \log(2)}{6615}
-\frac{120006708435844 \log^2(2)}{231525}-\frac{3071211711902343 \log(3)}{28028000}   \notag \\&
+\frac{1917380079}{175} \g_E  \log(3)-8089713 \pi ^2 \log(3)+\frac{6245376534}{175} \log(2) \log(3)
+\frac{1917380079 \log^2(3)}{350}  \notag \\&
+\frac{49326171875 \log(5)}{54432}  +\frac{88623623464 \zeta (3)}{2205}  \bigg) e^4
+ \cdots \biggr]  , \\
\label{eqn:B8L}
\mathcal{B}_{8L} =&  \frac{1}{(1-e^2)^{31/2}} \biggl[ -\frac{94860587410858}{3476347875}
-\frac{6486848 \g_E }{11025}-\frac{3434432 \pi ^2}{2205}+\frac{270364544 \log(2)}{77175}  \notag \\&
+ \bigg(  \frac{3471216185138587}{834323490}-\frac{2503138384 \g_E }{1323}+\frac{74170048 \pi ^2}{1323}
-\frac{97943219824 \log(2)}{46305}-1850202 \log(3) \bigg) e^2    \notag \\&
+ \bigg(  \frac{23616087848898761017}{83432349000}-\frac{3174674232748 \g_E }{33075}
+\frac{44311811732 \pi ^2}{6615}-\frac{28546607371664 \log(2)}{77175}   \notag \\&
+\frac{10573372989 \log(3)}{350} \bigg) e^4
+ \cdots \biggr] .
\end{align}

\section{PN expansion of the horizon angular momentum absorption to 18PN}
\label{sec:LfluxExps}

Prior results on the circular-orbit limit imply that the horizon angular momentum flux will have a series 
of the form
\begin{align}
\label{eqn:angMomFluxHor}
\left\langle \frac{dL}{dt} \right\rangle_H=  \frac{32}{5} \frac{\mu^2}{M} y^{15/2} &
\biggl[\mathcal{D}_0 + y\mathcal{D}_1 +y^2\mathcal{D}_2
+ y^3\biggl(\mathcal{D}_3+\mathcal{D}_{3L}\log y \biggr)
+ y^4\biggl(\mathcal{D}_4+\mathcal{D}_{4L}\log y \biggr)
\notag\\&\quad  + y^5\biggl(\mathcal{D}_5+\mathcal{D}_{5L}\log y \biggr)
+ y^{11/2}\mathcal{D}_{11/2} + y^6\biggl(\mathcal{D}_6
+\mathcal{D}_{6L}\log y +\mathcal{D}_{6L2}\log^2 y \biggr)
\notag\\&\quad+ y^{13/2}\mathcal{D}_{13/2}
 + y^7\biggl(\mathcal{D}_7
+\mathcal{D}_{7L}\log y +\mathcal{D}_{7L2}\log^2 y \biggr)+\cdots
\biggr] .
\end{align}
As with the absorbed energy, Forseth \cite{Fors16} used numeric-analytic fitting to find eccentricity coefficients 
in these angular momentum flux functions to 7PN horizon-relative order.  In particular, he found closed-form 
expressions for $\mathcal{D}_0(e)$, $\mathcal{D}_1(e)$, $\mathcal{D}_2(e)$, $\mathcal{D}_{3L}(e)$, 
$\mathcal{D}_{4L}(e)$, $\mathcal{D}_{6L2}(e)$.  He then extracted analytic finite-order series in $e^2$ for many of 
the remaining terms, specifically finding $\mathcal{D}_3(e)$ to $e^{40}$, $\mathcal{D}_4(e)$ to $e^{6}$, 
$\mathcal{D}_5(e)$ to $e^2$, $\mathcal{D}_{5L}(e)$ to $e^{32}$, $\mathcal{D}_{11/2}(e)$ to $e^{14}$, 
$\mathcal{D}_{6L}(e)$ to $e^4$, $\mathcal{D}_{13/2}(e)$ to $e^{4}$, and $\mathcal{D}_{7L2}(e)$ to $e^{6}$.  
No additional analytic coefficients were found in $\mathcal{D}_5(e)$, $\mathcal{D}_6(e)$, $\mathcal{D}_{7}(e)$, 
$\mathcal{D}_{7L}(e)$ beyond the known circular-orbit terms.  

Just as with the energy flux, we have extended the angular momentum absorption to $e^{20}$ through 10PN and $e^{10}$ 
through 18PN horizon-relative order, displaying a subset of the results to 8PN here.  The first three functions 
again yield closed forms \cite{Fors16}
\begin{align}
\mathcal{D}_0 =& \frac{1}{(1-e^2)^6}\biggl(1+\frac{15}{2} e^2+\frac{45}{8} e^4
+\frac{5 }{16}e^6\biggr),\\
\mathcal{D}_1 =& \frac{1}{(1-e^2)^7}\biggl(4+42e^2+\frac{15}{4}e^4-40e^6
-\frac{195}{64}e^8\biggr),\\
\mathcal{D}_2 =& \frac{1}{(1-e^2)^8}\biggl(-\frac{38}{7}+197e^2
+\frac{7965}{16}e^4+\frac{1175}{16}e^6+\frac{37825}{256}e^8
+\frac{495}{32}e^{10}\biggr) + \frac{30}{\sqrt{1-e^2}} \mathcal{D}_0.
\end{align}

The remaining flux terms exhibit the same patterns and structure as their energy flux counterparts.  We find 
closed-form expressions at 3PN and 3PN log, with the discussion surrounding \eqref{eqn:B3} and \eqref{eqn:B3L} 
pertaining
\begin{align}
\label{eqn:D3}
\mathcal{D}_3 =& \frac{1}{(1-e^2)^{9}}\biggl[ -\frac{633427}{4725}-\frac{1148221 e^2}{2100}-\frac{61667 e^4}{100}
-\frac{5046283 e^6}{720} - \frac{2070809 e^8}{640}   
-\frac{736891 e^{10}}{1536}-\frac{26905 e^{12}}{512}  \notag\\& 
+\sqrt{1-e^2} \left(\frac{8252956}{33075}+\frac{333023069 e^2}{132300}  
+\frac{48324481 e^4}{88200} -\frac{566970143 e^6}{1058400}+\frac{411843863 e^8}{264600}
+\frac{7425 e^{10}}{64}\right)  \biggr]  \notag\\&   -
\biggl[\frac{35}{107} \pi^2 + \log\left(\frac{1-e^2}{1+\sqrt{1-e^2}}\right) \biggr] \mathcal{D}_{3L} ,\\
\mathcal{D}_{3L} =& \frac{-1}{(1-e^2)^9}\biggl(\frac{1712}{105}
+\frac{15622}{35} e^2+\frac{9202}{5} e^4+\frac{3531 }{2}e^6
+\frac{749}{2} e^8+\frac{535}{64} e^{10}\biggr) .
\end{align}
The 4PN log term was known to be closed.  The 4PN non-log structure is similar to the description of \eqref{eqn:B4}
\begin{align}
\mathcal{D}_4 =& \frac{1}{(1-e^2)^{10}}\biggl[\frac{10859497}{22050}
+\frac{52}{3} \pi^2-\frac{1024}{15} \gamma_E-\frac{3980}{21}\log(2) +\bigg(\frac{3562043}{294} + 352 \pi^2 
- \frac{15296}{5}\gamma_\text{E}  \notag\\&  
 -\frac{ 108896}{35} \log(2) -\frac{20412}{5} \log(3)\bigg)e^2  + \bigg(\frac{1914399217}{58800} 
-239 \pi^2-\frac{109224 }{5} \gamma_E -\frac{ 18556441}{105} \log(2)   \notag\\&
+\frac{297432}{5} \log(3)\bigg)e^4  + \bigg(\frac{638653199}{5600} 
-6503 \pi^2  - \frac{206752}{5} \gamma_E +\frac{297014537}{135} \log(2)  -\frac{6446547}{16} \log(3)  \notag\\&
- \frac{330078125}{432} \log(5)\bigg)e^6  + \bigg( \frac{3562043}{294}-\frac{15296 \g_E }{5}
+352 \pi ^2-\frac{108896 \log(2)}{35}-\frac{20412 \log(3)}{5}  \bigg) e^8     \notag\\&
+ \bigg( \frac{131465564381}{1075200}-\frac{6363 \g_E }{2}-\frac{28273 \pi ^2}{16}
+\frac{8187206373887 \log(2)}{54000}+\frac{41897616724017 \log(3)}{512000}     \notag\\&
-\frac{3012572265625 \log(5)}{36864}-\frac{451516629783821 \log(7)}{13824000}  \bigg) e^{10}
+ \bigg( \frac{8149493677}{129024}-\frac{805 \g_E }{16}-\frac{2625 \pi ^2}{64}   \notag\\&
-\frac{2551616060639357 \log(2)}{1944000}-\frac{318863338438527 \log(3)}{512000}  
+\frac{359725572265625 \log(5)}{995328}     \notag\\&
+\frac{4314758300454931 \log(7)}{8294400} \bigg) e^{12}
+ \bigg( \frac{62126611699}{1505280}  +\frac{1119978423832346 \log(2)}{99225}   \notag\\&
+\frac{233938946375256303 \log(3)}{160563200}  -\frac{290951443134765625 \log(5)}{260112384}   \notag\\&
-\frac{3466444202976331129 \log(7)}{884736000}  \bigg) e^{14} +\cdots \biggr],\\
\mathcal{D}_{4L} =& \frac{1}{(1-e^2)^{10}}\biggl(-\frac{9148}{105}
-\frac{18240 }{7}e^2-\frac{356711}{35} e^4-\frac{3973}{5} e^6
+\frac{388523}{32} e^8+\frac{304913}{80} e^{10}+\frac{6415}{64} e^{12}
\biggr) .
\end{align}
From this point on, the angular momentum absorption terms continue to display structures that are parallel to 
those found in the horizon energy flux functions, with the descriptions surrounding equations \eqref{eqn:B5} 
through \eqref{eqn:B8L} being also relevant here
\begin{align}
\mathcal{D}_{5} =& \frac{1}{(1-e^2)^{11}}\biggl[\frac{2547493}{1372} + \frac{780}{7} \pi^2+\frac{2528}{35} \g_E 
-\frac{425612}{2205} \log(2) +\bigg(\frac{17459549369}{617400}+\frac{84776}{21} \pi^2 + \frac{75928}{5} \g_E 
\notag\\&  +\frac{2876344}{735} \log(2) + \frac{71442}{5} \log(3) \bigg)e^2 
+ \bigg(  -\frac{327317515241}{823200}+\frac{8364304 \g_E}{35}+\frac{602645 \pi ^2}{28}   \notag \\&
+\frac{961194533 \log(2)}{980}-\frac{777114 \log(3)}{5} \bigg) e^4
+ \bigg( -\frac{295832552489}{141120}+\frac{6758639 \g_E}{7}+\frac{76395 \pi ^2}{2}    \notag \\&
-\frac{60714697927 \log(2)}{5670}+\frac{869840613 \log(3)}{1120}+\frac{107205390625 \log(5)}{18144} \bigg) e^6
+\cdots \biggr],\\
\mathcal{D}_{5L} =&  \frac{1}{(1-e^2)^{11}}\biggl( \frac{21220}{49}+\frac{10568546 e^2}{735}
+\frac{105391081 e^4}{980}+\frac{9248095 e^6}{28}+\frac{49336999 e^8}{140}  \notag \\&
+\frac{33176401 e^{10}}{448}-\frac{22547523 e^{12}}{2560}-\frac{254885 e^{14}}{512}  \biggr)
+ \frac{45}{\sqrt{1-e^2}} \mathcal{D}_{3L},  \\
\mathcal{D}_{11/2} =&  \frac{\pi}{(1-e^2)^{23/2}}
\biggl(-\frac{109568}{1575}-\frac{2673716}{525} e^2-\frac{10478082}{175} e^4 - \frac{45144231221}{226800} e^6
-\frac{1487314873}{6720} e^8  \notag\\&
-\frac{59287955317343}{725760000} e^{10} - \frac{35598516307309 }{4354560000}e^{12}
-\frac{445860177201473}{4551966720000} e^{14} -\frac{7632479873521 e^{16}}{46820229120000}  \notag \\&
+\frac{97059177665259263 e^{18}}{1699012474306560000}
-\frac{50174510761076183 e^{20}}{4045267795968000000} + \cdots \biggr),  \\
\mathcal{D}_{6} =& \frac{1}{(1-e^2)^{12}}\biggl[ \frac{278408801583211}{48134047500}
-\frac{594376 \g_E }{2835}+\frac{17112988 \pi ^2}{33075}-\frac{256 \pi ^4}{45}  
-\frac{1485396268 \log(2)}{694575}   \notag \\&
-\frac{27392}{315} \pi ^2 \log(2) +\frac{1465472 \log^2(2)}{11025} -\frac{54784 \zeta (3)}{105} 
+ \bigg(  \frac{85821442460021}{2917215000} 
-\frac{6942352 \g_E }{189}   \notag \\&
+\frac{49728058 \pi ^2}{3675} 
-\frac{6488 \pi ^4}{15}-\frac{53846558282 \log(2)}{694575}-\frac{694216}{105} \pi ^2 \log(2)  
+\frac{37140556 \log^2(2)}{3675}   \notag \\&
-\frac{1689822 \log(3)}{35}  
-\frac{1388432 \zeta (3)}{35} \bigg) e^2  + \bigg(  -\frac{5240369975829997}{2333772000} 
-\frac{121960978 \g_E }{135}-\frac{113474836 \pi ^2}{2205}    \notag \\&
- \frac{15716 \pi ^4}{3}  
-\frac{2913769591493 \log(2)}{694575}-\frac{1681612}{21} \pi ^2 \log(2)   
+\frac{89966242 \log^2(2)}{735}  +\frac{230612589 \log(3)}{280} \notag \\&
-\frac{3363224 \zeta (3)}{7} \bigg) e^4 + \cdots \bigg] , \\
\mathcal{D}_{6L} =& \frac{1}{(1-e^2)^{12}}\biggl(
-\frac{1265945848}{694575}-\frac{16493890982 e^2}{231525}-\frac{104951231074 e^4}{231525}
-\frac{1678568830571 e^6}{926100}   \notag \\&
-\frac{2140195374283 e^8}{352800} - \frac{12294919487327 e^{10}}{1764000}
-\frac{6710991064951 e^{12}}{3528000} + \frac{311616563737 e^{14}}{3292800} \notag \\&
+\frac{21944030513653 e^{16}}{210739200}+\frac{1497519760589479 e^{18}}{22759833600}
+\frac{3625336614440057 e^{20}}{75866112000}
+\cdots \biggr)   \notag \\&
-2 \left[\frac{35}{107}\pi^2  + \log\left(\frac{1-e^2}{1+\sqrt{1-e^2}}\right)\right] \mathcal{D}_{6L2} ,\\
\mathcal{D}_{6L2} =& \frac{1}{(1-e^2)^{12}}\biggl(
\frac{1465472}{11025}+\frac{37140556}{3675} e^2+\frac{89966242}{735} e^4
+\frac{613632053}{1470} e^6+\frac{66484343}{140} e^8  \notag\\&\qquad
+\frac{1002371399}{5600} e^{10}  +  \frac{87664993 }{4800}e^{12}
+\frac{57245}{256} e^{14}
\biggr),\\
\mathcal{D}_{13/2} =& \frac{\pi}{(1-e^2)^{25/2}}\biggl( \frac{7239376}{33075}+\frac{498843526 e^2}{11025}
+\frac{32043307573 e^4}{33075}+\frac{10826350727947 e^6}{1905120}     \notag \\&\qq
+\frac{605479733577979 e^8}{50803200} +\frac{10827476512699331 e^{10}}{1128960000}
+\frac{256407320284306739 e^{12}}{91445760000}    \notag \\&\qq
+\frac{1675180844613905281 e^{14}}{7080837120000}    
+\frac{276193163158536266411  e^{16}}{110121178890240000}   \notag \\&\qq
-\frac{196594041130209959 e^{18}}{10194074845839360000}
-\frac{225861941667542473117  e^{20}}{1189308732014592000000}
 +\cdots \biggr),\\
\mathcal{D}_{7} =& \frac{1}{(1-e^2)^{13}}\biggl[ 
\frac{139830180452857}{4202178750}-\frac{41122541072 \g_E}{5457375}+\frac{876544 \g_E^2}{1575}
+\frac{1921663 \pi ^2}{945}-\frac{16384 \g_E \pi ^2}{45} \notag \\&
-\frac{121809908713 \log(2)}{5457375}+\frac{1753088}{525} \g_E\log(2)-\frac{44752}{45} \pi ^2 \log(2)
+\frac{7653496 \log^2(2)}{1575}-\frac{1458 \log(3)}{49}     \notag \\&   -\frac{784 \pi ^4}{45} 
-\frac{40352 \zeta (3)}{15} + \bigg( \frac{6197531723738801}{2941525125}-\frac{128020343932 \g_E}{165375}
+\frac{34055104 \g_E^2}{525}+\frac{94140496 \pi ^2}{1225}  \notag \\&
-\frac{636544 \g_E \pi ^2}{15}  
-\frac{2272 \pi ^4}{15}-\frac{11449289013128 \log(2)}{12733875}+\frac{125701888}{525} \g_E \log(2)
-\frac{1336928}{35} \pi ^2 \log(2)    \notag \\&
+\frac{544089008 \log^2(2)}{3675}-\frac{64972695078 \log(3)}{67375} 
+\frac{3744144}{25} \g_E \log(3)-\frac{244944}{5} \pi ^2 \log(3)    \notag \\&
+\frac{7488288}{25} \log(2) \log(3) +\frac{1872072 \log^2(3)}{25}-\frac{4942016 \zeta (3)}{35}  \bigg) e^2  
+\bigg( \frac{88742914458905034991}{2823864120000}    \notag \\&
-\frac{22844069195159 \g_E}{1819125} +\frac{625032368 \g_E^2}{525}+\frac{8425492576 \pi ^2}{11025}
-\frac{11682848 \g_E \pi ^2}{15} +\frac{41806 \pi ^4}{3}   \notag \\&
-\frac{2891722861514959 \log(2)}{38201625}+\frac{131570624}{9} \g \log(2)  
-\frac{1193860762}{315} \pi ^2 \log(2)  +\frac{40096841559 \log^2(2)}{1225}   \notag \\&
+\frac{39745130309019 \log(3)}{2156000}-\frac{460529712}{175} \g_E \log(3)  
+\frac{4304016}{5} \pi ^2 \log(3)-\frac{921059424}{175} \log(2) \log(3)  \notag \\&
-\frac{230264856 \log^2(3)}{175} -\frac{1240234375 \log(5)}{14112}-\frac{37047516 \zeta (3)}{35}  \bigg) e^4
+ \cdots \biggr] \\
\mathcal{D}_{7L} =& \frac{1}{(1-e^2)^{13}}\biggl[ -\frac{4017866767}{363825}+\frac{876544 \g_E}{525}
-\frac{20176 \pi ^2}{45}+\frac{1059728 \log(2)}{225} + \bigg( -\frac{596572565758}{848925}   \notag \\&
+\frac{34055104 \g_E}{175}-\frac{2471008 \pi ^2}{105} +\frac{869110496 \log(2)}{3675}
+\frac{5616216 \log(3)}{25}\bigg) e^2  + \bigg( -\frac{49420103890709}{5093550}    \notag \\&
+\frac{625032368 \g_E}{175}-\frac{6174586 \pi ^2}{35}+\frac{69443369578 \log(2)}{3675}
-\frac{690794568 \log(3)}{175}  \bigg) e^4 
+ \cdots \biggr]  , \\
\mathcal{D}_{7L2} =& \frac{1}{(1-e^2)^{13}}\biggl(\frac{1736824}{1575} + \frac{310988224}{3675} e^2
+\frac{284829399}{245} e^4 +\frac{3316579811}{735} e^6 +\frac{20100276671 e^8}{3920}   \notag \\&
+\frac{934711019 e^{10}}{1400} -\frac{53854863467 e^{12}}{67200}-\frac{94744327 e^{14}}{640}
-\frac{4834795 e^{16}}{2048} \biggr)  ,  \\
\mathcal{D}_{15/2} =& \frac{\pi}{(1-e^2)^{27/2}}\biggl( -\frac{284700044}{297675}-\frac{81036003964 e^2}{297675}
-\frac{64841337691 e^4}{7560}-\frac{1310374935437849 e^6}{17146080}     \notag \\&\qq
-\frac{1420721090175533561 e^8}{5486745600}-\frac{159166848358764247 e^{10}}{428652000}
-\frac{2988404293284553705799 e^{12}}{13168189440000}   \notag \\&\qq
-\frac{56478488808103952821241 e^{14}}{1032386052096000}
-\frac{18218890242335897733888553 e^{16}}{3964362440048640000}  \notag \\&\qq
-\frac{134485408556240559141476359 e^{18}}{321113357643939840000}
-\frac{7880069081437385949150751441 e^{20}}{36698669445021696000000}
 +\cdots \biggr) ,  \\
\mathcal{D}_{8} =& \frac{1}{(1-e^2)^{14}}\biggl[ \frac{57784184943753058541}{626368360117500}
+\frac{1439910584804 \g_E}{496621125}+\frac{711232 \g_E^2}{1225} +\frac{2185682224 \pi ^2}{231525}\notag \\&
+\frac{120448 \g_E \pi ^2}{315}-\frac{36992 \pi ^4}{315} -\frac{239862025427236 \log(2)}{10429043625}
+\frac{7649024 \g_E \log(2)}{6615} -\frac{6515456 \pi ^2 \log(2)}{6615}    \notag \\&
+\frac{24280192 \log^2(2)}{9261}+\frac{1458 \log(3)}{49}
-\frac{6868864 \zeta (3)}{735}  + \bigg( -\frac{1793079545089570389037}{417578906745000}   \notag \\&
+\frac{2171076650206094 \g_E}{496621125}-\frac{2348028208 \g_E^2}{11025}
+\frac{5048409139 \pi ^2}{17150}+\frac{5736928 \g_E \pi ^2}{21}   -\frac{2772704 \pi ^4}{315}   \notag \\&
+\frac{32280215575912501 \log(2)}{6952695750}
-\frac{230474816}{175} \g_E \log(2)+\frac{107401856}{735} \pi ^2 \log(2)   
-\frac{73082616704 \log^2(2)}{77175}     \notag \\&
+\frac{3376535894406 \log(3)}{875875}
-\frac{9255384}{25} \g_E \log(3)+\frac{1347192}{5} \pi ^2 \log(3)-\frac{29848176}{25} \log(2) \log(3) \notag \\&
-\frac{4627692 \log^2(3)}{25}+\frac{25289248 \zeta (3)}{735}  \bigg) e^2
+ \bigg( -\frac{1083282732353782898993897}{3340631253960000}
+\frac{43870606953871 \g_E}{294294}   \notag \\&
-\frac{32306192816 \g_E^2}{3675}-\frac{470504914363 \pi ^2}{102900}+\frac{327013984 \g_E \pi ^2}{35}
-\frac{1046995 \pi ^4}{7}   +\frac{8356953051}{245} \pi ^2 \log(2)   \notag \\&
+\frac{8118457969737885403 \log(2)}{13905391500}  -\frac{993501840704 \g_E \log(2)}{11025}
-\frac{94850785148611 \log^2(2)}{463050}   \notag \\&
-\frac{2406913973811 \log(3)}{52000}  +\frac{706033584}{175} \g_E \log(3)-\frac{20505312}{5} \pi ^2 \log(3)
+\frac{2900101968}{175} \log(2) \log(3)    \notag \\&
+\frac{353016792 \log^2(3)}{175}  +\frac{55662109375 \log(5)}{127008}+\frac{716949978 \zeta (3)}{49} \bigg) e^4
+\cdots \biggr]  , \\
\mathcal{D}_{8L} =& \frac{1}{(1-e^2)^{14}}\biggl[ 
-\frac{94860587410858}{3476347875}-\frac{6486848 \g_E}{11025}-\frac{3434432 \pi ^2}{2205}
+\frac{270364544 \log(2)}{77175}   \notag \\&
+ \bigg( \frac{489401285684009}{257507250}-\frac{11555797648 \g_E}{11025}  
+\frac{12644624 \pi ^2}{2205}-\frac{28608160928 \log(2)}{25725}-\frac{25220484 \log(3)}{25}  \bigg) e^2 \notag \\&
+ \bigg( \frac{127638001836320407}{1158782625}-\frac{27455536336 \g_E}{735} 
+\frac{119491663 \pi ^2}{49} -\frac{11532637472737 \log(2)}{77175}    \notag \\&
+\frac{2547085176 \log(3)}{175}  \bigg) e^4
+\cdots \biggr] . 
\end{align}
Finally, as we mentioned in the discussion of \eqref{eqn:B8L2}, the 8PN $\log^2$ angular momentum term did settle 
into a resummed closed-form function
\begin{align}
\label{eqn:D8L2}
\mathcal{D}_{8L2} =& \frac{1}{(1-e^2)^{14}}\biggl( 
-\frac{49603088}{8575}-\frac{22345793836 e^2}{25725}-\frac{201203215223 e^4}{10290}
-\frac{1196416933477  e^6}{8575}    \notag \\&
-\frac{4391775215885 e^8}{10976}-\frac{598457874498 e^{10}}{1225}-\frac{76198202080153  e^{12}}{313600}-
\frac{9321366120429 e^{14}}{219520}  \notag \\&
-\frac{18037498466101 e^{16}}{10035200}-\frac{32570455 e^{18}}{8192} \biggr) 
+ \frac{60}{\sqrt{1-e^2}} \mathcal{D}_{6L2}  .
\end{align}

\end{widetext}

\section{Discussion}
\label{sec:Disc}

The results presented in the previous two sections, and at the online repositories \cite{BHPTK18,UNCGrav22}, have 
pushed the knowledge of the black hole horizon absorption in eccentric-orbit nonspinning EMRIs to 10PN (in an 
$e^{20}$ expansion) and to 18PN (in an $e^{10}$ expansion) relative to the leading horizon contribution.  Between our 
new fully symbolic calculations and earlier numerical high-precision fitting \cite{Fors16}, we have been able to 
discover closed-form eccentricity dependence for a host of terms: $\mathcal{B}_{0}$, $\mathcal{B}_{1}$, 
$\mathcal{B}_{2}$, $\mathcal{B}_{3}$, $\mathcal{B}_{3L}$, $\mathcal{B}_{4L}$, $\mathcal{B}_{5L}$, $\mathcal{B}_{6L2}$, 
$\mathcal{B}_{7L2}$, $\mathcal{D}_{0}$, $\mathcal{D}_{1}$, $\mathcal{D}_{2}$, $\mathcal{D}_{3}$, $\mathcal{D}_{3L}$, 
$\mathcal{D}_{4L}$, $\mathcal{D}_{5L}$, $\mathcal{D}_{6L2}$, $\mathcal{D}_{7L2}$, $\mathcal{D}_{8L2}$.  
Closed forms can likely also be found for $\mathcal{B}_{8L2}$ and $\mathcal{D}_{9L3}$, but our present 
expansions stop just short of providing confirmation that the series are finite.  The other terms up to 10PN 
horizon-relative order are apparent infinite series and our eccentricity expansions go deep enough to reveal 
structures resembling those seen in the infinity-side fluxes.

Particularly of note are the log sequences that appear in the infinity-side fluxes, which we defined and discussed 
previously \cite{MunnEvan19a,MunnEvan20a,MunnETC20}.  The leading-log (also called 0PN log) sequence, for example, 
starts with the Peters-Mathews term, $\mathcal{L}_0(e)$, includes the first appearance of a log, at 
$\mathcal{L}_{3L}(e)$, and continues with each new power of log at 6PN, 9PN, etc.  In other words, the terms in 
this sequence have PN orders $y^{3k} \, \log^k y$ ($k\ge 0$).  There is a companion half-integer-order leading log 
sequence \cite{MunnEvan19a} that is made up of the terms $y^{3k+3/2} \, \log^k y$ ($k\ge 0$), which starts 
($\log^0$) with the 1.5PN tail.  There are also integer and half-integer 1PN \cite{MunnEvan20a}, 2PN, 
3PN \cite{MunnEvan19a}, 4PN \cite{MunnEvan20a}, log sequences.  

In the horizon fluxes, the same set of integer-order log sequences appear and half-integer-order logs show up as 
well, with an important caveat (discussed below).  Our results show that, just as was found in the fluxes 
at infinity, the leading-, 1PN-, and 2PN-log series have purely rational number coefficients, with the first 
appearance of transcendental numbers occurring in the 3PN log sequence.  As \cite{MunnEvan19a,MunnEvan20a} showed, 
the presence of only rational number coefficients indicates that these terms arise merely from low-multipole-order 
source moments.  Furthermore, the 2PN logarithms once again display a dominant-subdominant eccentricity singular 
factor structure, with the subdominant term being proportional to the corresponding leading-log flux (see e.g., 
\eqref{eqn:B5L} with the appearance of \eqref{eqn:B3L}).  

A significant difference, however, is the delay in the appearance of half-integer-order flux terms in the horizon 
absorption.  On the infinity-side, the first half-integer contribution is the tail term at 1.5PN order.  In the 
horizon fluxes, the first appearance of a half-integer term is at 5.5PN order (which, of course, is at 9.5PN 
relative to the leading flux at infinity).  The tail contribution at infinity stems from a nonlinear interaction 
between the outgoing (Newtonian) quadrupole wave and the static mass monopole \cite{BlanScha93,ArunETC08a}.  It 
appears that the combination of weak backscatter toward the primary black hole and the small cross section leads 
to an 8PN suppression of the tail flux at the horizon. 

The horizon flux terms at 5.5PN, 6.5PN, and 7.5PN horizon-relative order all involve rational number series (once 
an overall factor of $\pi$ is pulled out).  If the infinity-side fluxes are any guide, it may be that these 
half-integer leading-log, 1PN-log, and 2PN-log sequences, respectively, can be linked to the 0PN, 1PN, and 2PN 
horizon-relative flux terms.  Stated another way, we showed in \cite{MunnEvan19a} and \cite{MunnEvan20a} that all 
leading logarithms (integer and half-integer) are determined completely by certain sums over the Newtonian mass 
quadrupole power spectrum $g(n,e)$, which completely determines the dominant Peters-Mathews flux, $\mathcal{L}_0(e)$.  
Similarly, all terms in the 1PN-log sequence are determined solely by the power spectra of the 1PN multipoles (i.e., 
the Newtonian mass octupole, Newtonian current quadrupole, and 1PN correction to the mass quadrupole), which are 
the sole ingredients that determine the 1PN flux, $\mathcal{L}_1(e)$.  It is possible that a FD multipole 
formulation of the horizon fluxes could show similar linkage between leading integer and half-integer logarithms.  
A multipole formulation might also lead to horizon-flux analogs of the 3PN enhancement functions 
$\chi(e)$ and $\tilde{\chi}(e)$ \cite{MunnEvan20a}, which could aid in finding compact forms for complicated 
functions like $\mathcal{B}_4$.  On the infinity-side, the function $\chi(e)$ 
shows up in the 3PN non-log flux.  It is interesting to note that a comparable infinite series does not appear in 
the 3PN non-log horizon flux, nor is there an appearance of the Euler gamma constant, $\g_E$.  Lack of these terms 
greatly facilitated the process of extracting the closed forms for $\mathcal{B}_{3}$ and $\mathcal{D}_3$ found in 
\eqref{eqn:B3} and \eqref{eqn:D3}, respectively.  

\begin{figure*}
\hspace{-1.5em}\includegraphics[scale=.71]{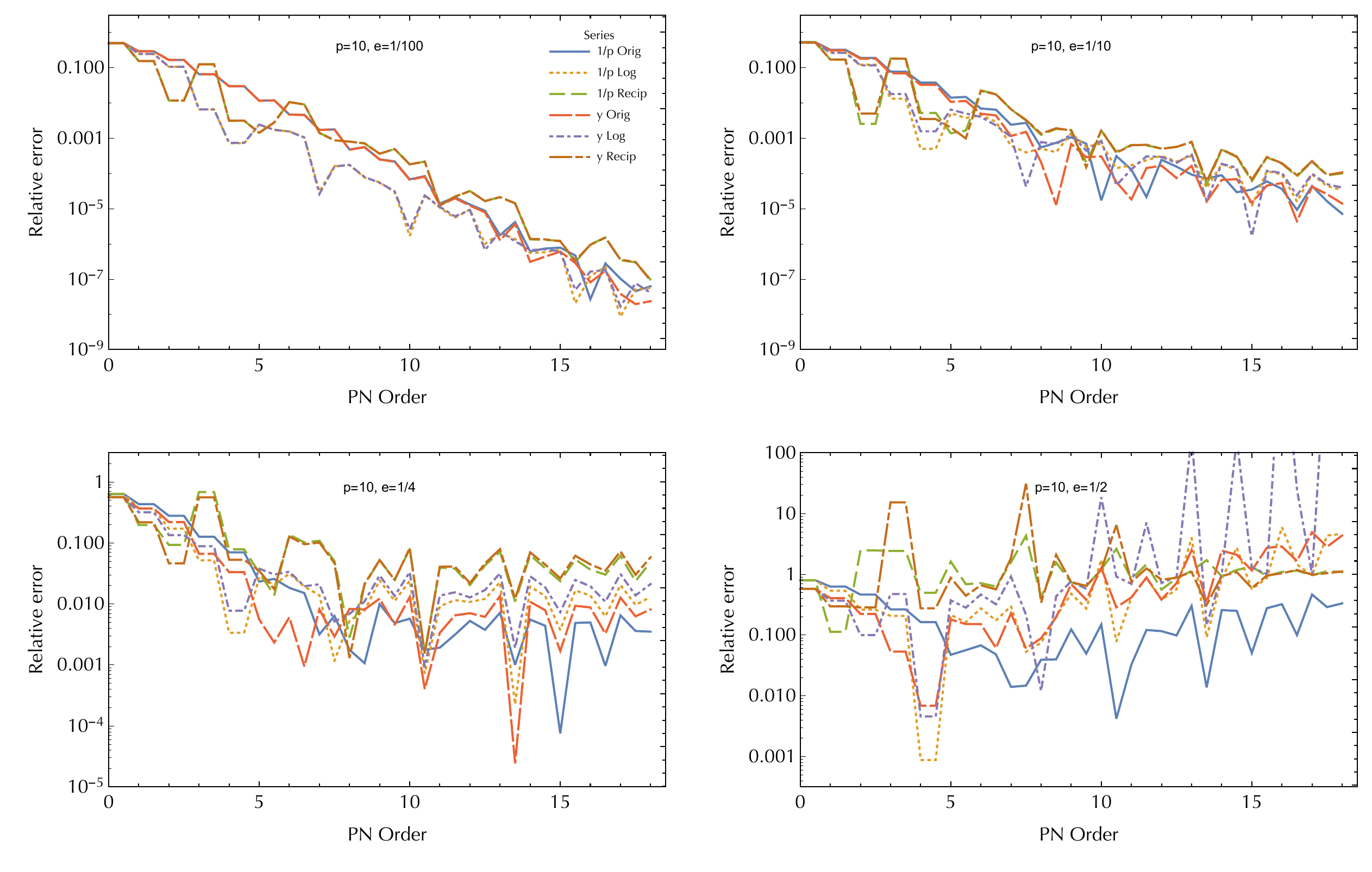}
\caption{Accuracy of the tidal heating PN expansion and several resummations for a set of close orbits.  An 
orbital separation of $p=10$ is chosen.  Four different eccentricities are tested: $e=1/100$, $e=1/10$, $e=1/4$, 
and $e=1/2$ (each a separate panel).  The numerical values generated by inserting the orbital parameters into 
our PN expansion are plotted against accurate numerical flux values obtained from a Teukolsky code for these orbits.
Each plot shows the relative error as a function of stepping toward higher PN order.  The multiple curves on each 
plot follow from using PN expansions computed with different compactness parameters (i.e., $1/p$ and $y$) and 
with and without the use of logarithmic and reciprocal resummations.  The asymptotic PN series for the horizon 
energy flux breaks down for an orbit as tight as $p = 10$ when the eccentricity is higher than $e = 0.25$.  However, 
at $e=0.25$ and lower, useful accuracy is available in the expansions if summed through a number of PN orders.
\label{fig:p10ords}}
\end{figure*}

\section{Testing convergence of the horizon flux PN expansions on a close orbit}
\label{sec:numComps}

We now use the expansions laid out in the preceding sections to make comparisons with numerical horizon flux data 
to assess their accuracy and convergence.  We confine the presentation here to the energy flux case but note that 
the angular momentum expansion yields similar results.  To prepare for the numerical comparison, we assemble a 
net expansion by combining flux terms from three sources.  First, some of the PN terms are closed-form expressions 
in $e$ and these provide exact inputs.  Second, in cases of flux terms that are not closed-form, out to 10PN 
horizon-relative order we use the full $e^{20}$ eccentricity expansions.  Finally, beyond 10PN out to 18PN, we 
use the full $e^{10}$ eccentricity expansions.  A separate Teukolsky code was used (J.~Castillo, private 
communication) and specialized to $a=0$.  Horizon flux data were generated for $p=10$ separation orbits with 
eccentricities of $e = \{1/100, 1/10, 1/4, 1/2\}$ to mirror the similar comparisons made in \cite{Munn20} for the 
energy flux at infinity.  The results are given in Fig.~\ref{fig:p10ords}.

When we compare to \cite{Munn20}, we find that the horizon flux expansions exhibit worse fidelity for $p=10$ than 
their counterpart infinity-side fluxes, particularly as $e$ grows large.  Whereas the flux at infinity demonstrated 
steady average reduced error with increasing PN order all the way to $e=1/2$, the horizon flux breaks down beyond 
$e=1/4$.  At $e=1/4$, the error decreases roughly monotonically until 8PN, after which it begins to fluctuate or 
grow (typical of an asymptotic series).  The evaluation at $e=1/2$ is worse, disconnected from any convergent 
behavior at and beyond 7PN order.  At $e=1/2$, the expansion only briefly exhibits an error less than $1\%$.  
It is likely that this is due, in part, to the fact that the eccentricity enhancement functions in the expansion 
take on increasingly large numerical values as $e \rightarrow 1$ in the expansion that uses $y$ as the compactness 
parameter.  Lending support to this claim is that the expansion in $1/p$ (blue curve) is better behaved.  

It may be that taking the eccentricity expansions only out to $e^{20}$ is insufficient for generating accurate 
values at $e \ge 1/2$ on $p = 10$ orbits.  One way to see this is to compare how the frequency-domain Teukolsky 
code generates accurate values for high-eccentricity orbits.  In that code, $lmn$ modes are computed, with $n$ 
being the harmonics of the radial libration frequency.  At high eccentricity, increasingly large numbers of $n$ 
modes are required to reach, for example, 12 digits of accuracy in the flux.  The fully symbolic PN expansion 
code is similar, where the PN expansions must be built out of MST modes and where we are required to compute 
$n$ modes with $|n|$ up to half the desired maximum eccentricity order.  In other words, if the Teukolsky code is 
using a greater number of modes at an eccentricity of $e=1/2$ than our symbolic code is, it may indicate we need 
to take the $e$ expansion to higher order.  Insufficient mode representation has also been noted as a limiting factor 
for small $p$ \cite{Fuji12b, Fuji15}.  Nevertheless, it is encouraging that the accuracy of the 
infinity-side full-flux expansion was fairly strong even for the $p=10, e=1/2$ orbit, owing to the use of 
arbitrary-order eccentricity expansions at low PN and the use of eccentricity resummations throughout, and for 
$e\le 1/4$ our horizon fluxes could be usefully added to produce a net energy loss.

\section{Conclusions}
\label{sec:horConcs}

This paper has described new high-PN-order results for the tidal heating and torquing (also referred to as horizon 
fluxes or horizon absorption) onto a nonspinning primary black hole in an eccentric-orbit EMRI.  The present work 
extends earlier calculations \cite{BiniDamo13,KavaOtteWard15,Fuji15} to orbits with eccentricity and gives analytic 
expressions for the fluxes in two expansions: one to 10PN horizon-relative order in an $e^{20}$ eccentricity 
expansion and the other to 18PN in an $e^{10}$ expansion.  These calculations represent a significant extension over 
previous work with numeric-analytic fitting, which was available only in Forseth's thesis \cite{Fors16} and several 
online talks \cite{Fors16b,Evan16}.  Taken together with high-order expansions of the infinity-side fluxes 
\cite{Munn20}, the full dissipation in eccentric-orbit nonspinning EMRIs is now known to 19PN order.

Several remarkable features exist in the form of the horizon absorption expansions, especially the presence at low 
PN order of a number of closed-form-in-$e$ terms with simple rational number coefficients and the delayed appearance 
(to 5.5PN horizon-relative order) of the first half-integer (likely tail) term.  The combined structure suggests 
that a focused calculation using post-Newtonian theory might allow some of these low-order terms to be 
calculated directly, rather than extracted from first-order black hole perturbation theory, as was possible with 
certain PN-log sequences in the flux to infinity \cite{MunnEvan19a,MunnEvan20a}.  Of course, a direct PN calculation 
must deal with the fact that the primary black hole horizon and nearby region do not naturally and immediately fit 
within PN theory.

We also tested numerically the convergence (i.e., in the sense of an asymptotic series) of the tidal heating PN 
expansion when extended to close orbits with $p=10$.  We found the results to be less convergent than was the case 
with the infinity-side flux \cite{Munn20}, especially for eccentricities approaching $e\sim 1/4$ and higher.  
However, at $e = 1/4$ and $p=10$ it is possible to achieve a calculation of the flux with less than 1\% error using 
the full PN expansion, and even $e = 1/2$ at $p=10$ can yield a 90\% accurate result with the best-case resummation 
of the series.  When we account for the fact that the tidal heating is suppressed by 4PN in the nonspinning primary 
case, the fractional errors in using the PN expansion for this size orbit would be of order $10^{-5}$ or less in the 
total dissipation.  Use of the PN expansion improves rapidly with increased orbital separations.

Our fully symbolic calculations used \textsc{Mathematica} in parallel on a cluster computer.  As with the expansion 
at infinity \cite{Munn20}, the bottleneck step in the procedure was the calculation of the even-parity asymptotic 
amplitudes for $l=2$.  Part of the calculation is sequential and part can be made parallel, by splitting over 
modes.  To give a sense of the speed of the code, our 18PN horizon-relative calculations on the UNC cluster 
(Longleaf) were measured in days.  The attempt to reach 19PN order failed to complete in under 10 days.  

\acknowledgments

We thank Jezreel Castillo for running his Teukolsky code to provide numerical data that we used in 
Sec.~\ref{sec:numComps} to test convergence of our PN expansions.  This work was supported by 
NSF Grant Nos.~PHY-1806447 and PHY-2110335 to the University of North Carolina--Chapel Hill.  C.M.M.~acknowledges 
additional support from NASA ATP Grant 80NSSC18K1091 to MIT.

\clearpage

\bibliography{horizon}

\end{document}